 \documentclass[preprint2,longabstract]{aastex}

\slugcomment{Not to appear in Nonlearned J., 45.}

\shorttitle{A model for $\gamma$-ray emission in radio galaxies}
\shortauthors{P. Banasi\'nski \& W. Bednarek}

\begin{document}


\title{Two blobs in a jet model for the $\gamma$-ray emission in radio galaxies}


\author{P. Banasi\'nski \& W. Bednarek}
\affil{Department of Astrophysics, The University of Lodz, 90-236 Lodz, ul. Pomorska 149/153, Poland}

\email{p.banasinski@uni.lodz.pl, bednar@uni.lodz.pl}



\begin{abstract}
In the unified scheme FR I type radio galaxies are identified with the  blazar type active galaxies which jets are aligned at large angles to the line of sight.  A few radio galaxies of this type have been discovered to emit GeV-TeV gamma-rays. We consider a scenario which naturally explains the very high energy gamma-ray emission at large angles to the jet axis. It is proposed that two emission regions are present in the jet at this same moment. The inner region (blob I), moves with the large Lorentz factor, producing radiation strongly collimated along the jet axis, as observed in BL Lac type blazars. On the other hand, the outer region (blob II), which moves with the mild Lorentz factor, contains isotropically distributed relativistic electrons in the blob reference frame. These electrons up-scatter mono-directional soft radiation from the blob I preferentially in the direction opposite to the jet motion. Therefore, gamma-rays, produced in the blob II, can be emitted at a relatively large angles to the jet axis in the observer's reference frame. We analyze the basic emission features of such external blob radiation model. The example modeling of the emission from the FR I type radio galaxy, NGC 1275, is presented.  
\end{abstract}




%
%
\section{Introduction}

According to the unification scheme, FR I type active galactic nuclei (AGNs) are interpreted as BL Lac type blazars viewed at large angles to the jet axis. Such interpretation is supported by the observations of the inner radio jets in these sources which are oriented at large angles to the line of sight, e.g. in Cen A $i\sim (12-45)^\circ$ (M\"uller et al.~2014) and $i\sim (50-80)^\circ$ (Tingay et al.~2001), in NGC 1275 $i\sim (30 - 55)^\circ$ (Walker et al.~1994, Vermeulen et al.~1994) and $i\sim (65\pm 16)^\circ$ (Fujita \& Nagai~2017), and in M87 $i\sim (10-19)^\circ$ (Biretta et al.~1999). A few nearby FR I type radio galaxies, M87, Cen A, and NGC 1275, have been discovered to emit $\gamma$-rays extending up to TeV energies. In the case of M87, the differential TeV $\gamma$-ray spectrum is flat, well described by the power law with the index 2.2$\pm$ 0.2 in the high state. It extends up to $\sim$10 TeV (Aharonian et al.~2003, Albert et al.~2008, Acciari et al.~2009, Aliu et al.~2012). Moreover, the TeV emission is observed to vary on a time scale of a day. The spectrum in the ${\it Fermi}$-LAT energy range is complex showing hardening above $\sim$10 GeV (Abdo et al.~2009a). The GeV emission below the break shows variability on a month time scale (Ait Benkhali et al.~2018) and a day to a week time scale  (Tanada et al.~2018). Such a short time scale variability of the low energy GeV component is also observed in the case of another radio galaxy Cen A (Sahakyan et al.~2013, Brown et al.~2017). The TeV $\gamma$-ray emission from Cen A has been detected up to $\sim$5 TeV showing also relatively flat spectrum (spectral index $-2.7\pm 0.5$, Aharonian et al.~2009). On the other hand, the 
$\gamma$-ray emission from NGC 1275 shows different behavior. It is observed only up to $\sim$600 GeV with the steep spectrum, spectral index $\sim$4.1 (Aleksi\'c et al.~2012, 2014a). The {\it Fermi}-LAT emission (discovered by Abdo et al.~2009b, Kataoka et al.~2010) shows strong variability up to an hour time scale (Brown \& Adams~2011, Baghmanyan et al.~2017). 
The GeV emission has steadily increased over 8-year time-span of the {\it Fermi}-LAT observations (Tanada et al.~2018).
Recently, two very large flares from NGC 1275 has been observed in the GeV energies (Pivato \& Buson~2015, Lucarelli et al.~2017) and TeV energies (Mirzoyan~2017, Mukherjee 2017, Ansoldi et al.~2018), with the flux increase by a factor of $\sim$60.
  
The short time scale variability, of the order of days to hours observed at GeV-TeV energies, indicates that the emission in FR I type radio galaxies originates in the inner part of the jet or in the direct vicinity of the super-massive black 
hole (SMBH), see for review Rieger~(2017). In the standard blob in a jet model the emission region should move rather mildly relativistic in order to produce $\gamma$-rays at the angles of the order of $\sim$40$^\circ$-$50^\circ$ to the jet axis, as estimated in the case of Cen A and NGC 1275. Such emission might be explained in terms of the mildly relativistic homogeneous synchrotron self-Compton model (e.g. Chiaberge et al.~2001) or in terms of a more complicated structured jets (Ghiselini \& Tavecchio~2008, Giannios et al.~2009). In fact, the high energy $\gamma$-ray emission at large angles to the jet axis is also expected from the region surrounding the inner jet in the inverse Compton (IC) $e^\pm$ pair cascade models (e.g. Sitarek \& Bednarek~2010, Roustazadeh \& B\"ottcher~2011). The emission, varying on a short time scale at large angle to the jet axis, has been also expected in terms of the star/cloud-jet collision model in which the jet plasma interacts directly with the wind of the massive star (Bednarek \& Protheroe~1997) or with the matter of the red giant (e.g. Barkov et al.~2010). The last class of models link large angle, variable $\gamma$-ray emission to the processes occurring in the magnetosphere of the rotating SMBH (e.g. Rieger \& Mannheim~2002, Neronov \& Aharonian~2007, Rieger \& Aharonian~2008, Levinson \& Rieger~2011, Aleksi\'c et al.~2014b). It is not clear at present whether any of the above mentioned models can explain complex features of the $\gamma$-ray emission from radio galaxies. 
 
Here we consider the standard blob in a jet scenario for the variable, high energy emission from the inner jet in radio galaxies. However, it is assumed that two emission regions are present within the jet at this same moment. The inner region (called blob I) moves fast and emits highly collimated radiation towards the outer, slowly moving region (called blob II).
Isotropically distributed relativistic electrons, in the blob II reference frame, up-scatter collimated radiation from the inner blob I, preferentially in the direction opposite to the jet propagation. Therefore, 
$\gamma$-rays, produced in the blob II, are emitted at larger angles to the observer's line of sight than expected in the case of the isotropic emission region. 
On the other hand, radiation produced in blob I is highly collimated along the jet axis. This emission is characteristic for the BL Lac type objects observed at a small viewing angles. We discuss the basic emission features of such external blob radiation model. We argue that the model can describe the high energy emission from radio galaxies viewed at large angles. 
Note that our model differs essentially from that proposed by Georganopoulos et al.~(2005) in which 
radiation from the slow blob II is up-scattered by the relativistic electrons in the fast blob I. 
That model seems to be more suitable for the AGNs observed at small inclination angles, i.e. BL Lacs.
On the other hand, our scenario shows similarities to the external Compton disk model (Dermer, Schlickeiser \& Mastichiadis~1992) in which the accretion disk radiation is scattered by electrons in the blob moving relativistically along the jet. This model has been studied in a more detail in Dermer \& Schlickeiser~(1993) and Dermer, Sturner \& Schlickeiser (1997).

\begin{figure}
\vskip 3.5truecm
\includegraphics{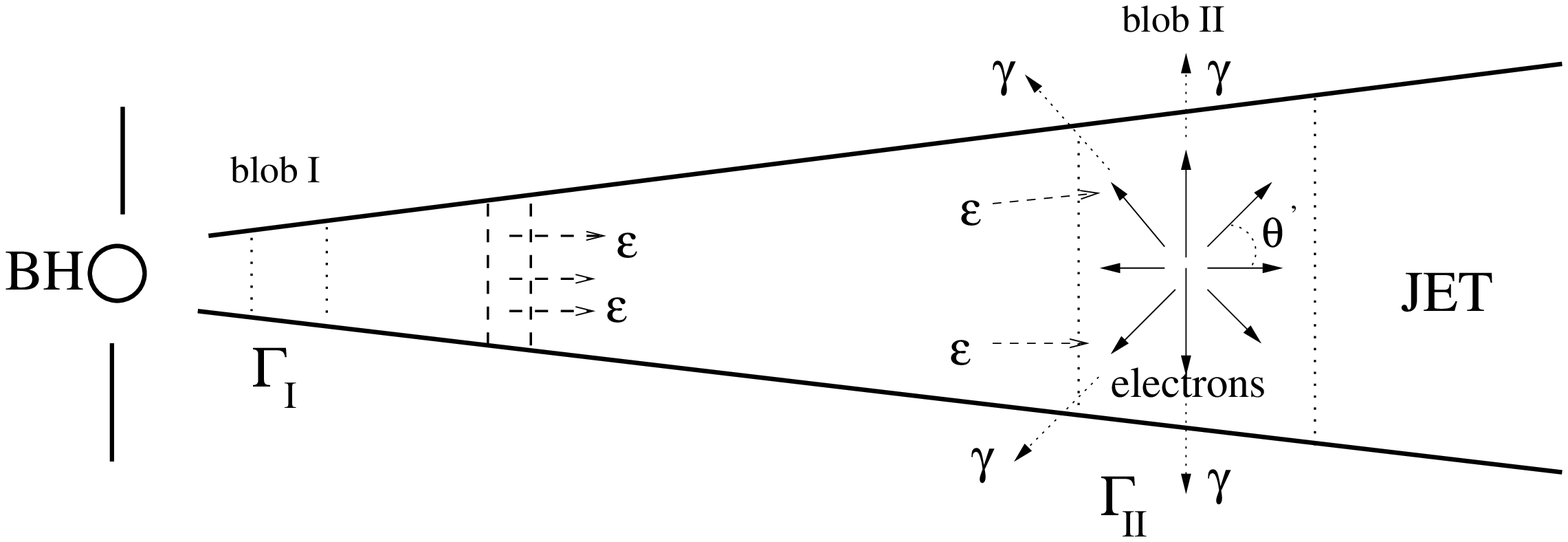}
\caption{Schematic representation of the jet in a radio galaxy with two blobs (marked by 'blob I' and 'blob II') moving with the Lorentz factors of $\Gamma_{\rm I}$ and $\Gamma_{\rm II}\ll \Gamma_{\rm I}$. The blob I (closer to the jet base) emits strong soft radiation, $\varepsilon$, which is Doppler busted towards the blob II (at larger distance from the jet base). The relativistic electrons, in the blob II, comptonize the soft photons from the blob I to the $\gamma$-ray energy range. The blob I moves with the large Lorentz factor and the blob II moves with a relatively small Lorentz factor. Therefore, IC $\gamma$-rays ($\gamma$) are produced at a relatively large angle, $\theta '$, to the jet axis measured in the blob reference frame.}
\label{fig1}
\end{figure}
\section{External blob radiation model}

Jets in active galaxies are expected to have complex structure. Their inner parts, close to the jet base, are composed of plasma which moves with large Lorentz factors. The emission from this region is relativistically boosted along the jet axis. The plasma in the outer parts, farther from the jet base, is decelerated as a result of interaction with the matter within the jet. We consider two separated regions in the jet, called blob I and blob II. Both blobs contain relativistic electrons and emit radiation in the synchrotron and the IC processes.    
In terms of such two blobs scenario the emission from the BL Lac type blazars is identified with 
the emission from the inner blob I. In fact, the TeV $\gamma$-ray emission from two extreme BL Lac type objects, Mrk 501 and PKS 2155-304 is strongly variable on a time scale down to a few minutes (Albert et al.~2007, Aharonian et al.~2007). Such short variability requires the presence of the emission region in the jet which moves with the Lorentz factor over 50 (Begelman et al.~2008). On the other hand, observations of the superluminal motion of radio emitting regions in the parsec scale jets suggest much lower values of the jet Lorentz factors. It is clear that the jet has to decelerate at some distance after initial propagation with the large Lorentz factor. If this deceleration occurs already in the inner jet, then this region will produce radiation at a relatively large angle to the jet axis. We show that the large angle emission can be significantly enhanced if the relativistic electrons in the blob II inverse Compton up-scatter mono-directional soft radiation produced in the blob I (see the geometrical situation presented in Fig.~1).
Therefore, the existence of such two regions in the jet, either well localized or quite extended, seems to provide natural mechanism for the $\gamma$-ray emission observed from radio galaxies. Such geometrical structure of the jet, the inner fast blob and the outer slow blob, creates the baseline for the model discussed here. Note that such general structure of the jet has been already discussed by Georganopoulos et al.(2005). However, in their model the high energy emission is produced in the blob I as a result of up-scattering of the soft radiation produced in the blob II. We consider the opposite scenario in which the high energy emission is produced in the blob II as a result of up-scattering the mono-directional soft radiation arriving from the blob I. Due to the geometrical effects of the IC scattering process of the mono-directional radiation (from the blob I) by isotropic relativistic electrons which are isotropic in the blob II reference frame, the $\gamma$-ray emission from blob II can be produced at large angles to the jet axis. We argue that the emission from the blob II is responsible for the $\gamma$-ray emission observed from FR I type radio galaxies observed at the angle of a few tens degrees to the jet axis.   

In the considered two blob in a jet scenario, both blobs can in fact follow this same pattern during the propagation in the jet. The initially very relativistic jet can significantly decelerate as a result of the interaction with the matter surrounding the SMBH. Then, the Lorentz factor of the jet drops with the propagation distance according to the prescription expected in the case of the relativistic jets in Gamma-Ray Bursts, i.e. 
$\Gamma_{\rm j}(z)\propto \Gamma_{\rm j,0}(z/z_0)^{-3/2}$ (see Blandford \& McKee~1976 and Sari~1997), where 
$z$ is the distance measured from the base of the jet and
$\Gamma_{\rm j,0}$ is the Lorentz factor of the jet at the distance $z_0$ from its base. The jet plasma  
with the initial Lorentz factor of the order of $\Gamma_{\rm j}(z_0) = 100$, e.g. at the distance $z_0 = 3\times 10^{15}$~cm from the base of the jet, can decelerate to the Lorentz factor of the order of $\sim$3 at the propagation distance of $\sim$10 R$_0  = 3\times 10^{16}$~cm. The above mentioned example parameters postulate that the acceleration of the blobs in the jet occurs already within a few tens of the Schwarzschild radii of a SMBH with the mass of the order of $3\times 10^8$~M$_\odot$.

In order to perform the calculations of the $\gamma$-ray spectra produced in such external blob radiation model, we have to determine the soft radiation field produced in the blob I. Note that the radiation from the blob I is strongly collimated along the jet axis, so it cannot be observed directly in the case of a specific FR I type radio galaxy which is observed at large angle to the jet axis. Therefore, for the purpose of these studies we apply the results of observations of the soft emission typical for the classical BL Lacs which are viewed at small angles, e.g. Mrk 501. As an example, we assume that the flux of soft radiation, produced in the blob I but observed at the distance to the Earth, is approximated by a simple power law function,
\begin{eqnarray}
F(\varepsilon) = B\varepsilon^{-\alpha}~~~{\rm ph.~cm^{-2}~s^{-1}~MeV^{-1}}, 
\label{eq1}
\end{eqnarray}
\noindent
below the maximum energy $\varepsilon_{\rm max}$ and "B" is the normalization constant. 
Then, the density of photons from the blob I, which is measured in the blob II reference frame, is 
\begin{eqnarray}
n(\varepsilon) = F(\varepsilon)d_{\rm L}^2D_{\rm II}^2/(z^2c)~~~{\rm ph. cm^{-3}~MeV^{-1}}, 
\label{eq2}
\end{eqnarray}
\noindent
where $d_{\rm L}$ is the luminosity distance to the specific BL Lac, and $D_{\rm II} = [\Gamma_{\rm II}(1 + \beta_{\rm II})]^{-1}$ is the Doppler factor of the blob II applied for the transformation of the soft radiation field from the observer's reference frame to the blob II reference frame, $\Gamma_{\rm II}$ is the Lorentz factor of the blob II, its velocity $\beta_{\rm II} = (\Gamma_{\rm II}^2 - 1)^{0.5}/\Gamma_{\rm II}$, and $c$ is the speed of light. 
Note that radiation produced in the blob I approaches from behind the blob II (see Fig.~1).
Above defined radiation field can be seen only by the observer located at the jet axis, i.e. such a source belongs to the class of the BL Lacs according to the classification scheme. Therefore, in order to perform some realistic example calculations, the parameters of the emission stage observed in the case of a typical BL Lac type source such as Mrk 501, should be used (see Pian et al.~1998, and also Eq.~2 in Bednarek \& Protheroe~1999). 

\begin{figure}
\vskip 5.5truecm
\includegraphics{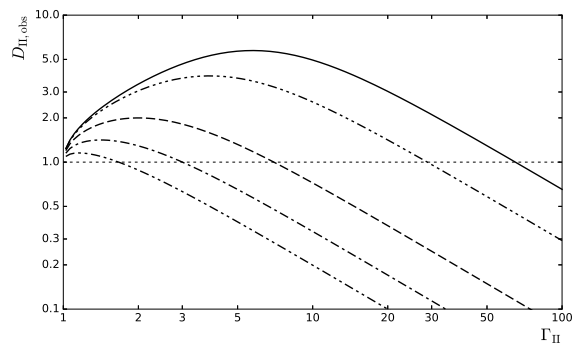}
\caption{The Doppler factor of the blob II, $D_{\rm II,obs}$, as a function of its Lorentz factor for selected observation angles measured in respect to the jet axis, $\theta = 10^\circ$ (solid), $15^\circ$ (dot-dot-dot-dashed), $30^\circ$ (dashed), $45^\circ$ (dot-dashed), and $60^\circ$ (dot-dot-dashed). }
\label{fig2}
\end{figure}

We calculate the $\gamma$-ray spectra produced by isotropic electrons with the equilibrium spectrum, 
$dN_{\rm e}/dE_{\rm e}$, which scatter mono-directional soft radiation with the spectrum, $n(\varepsilon)$,
in the blob II reference frame, by integration the following formula,
\begin{eqnarray} 
{{dN}\over{d\varepsilon_\gamma dt'd\Omega'}} = \int_{E_{\rm L}}^{E_{\rm U}} {{dN_{\rm e}}\over{dE_{\rm e}}}\int_{\varepsilon_{\rm L}}^{\varepsilon_{\rm U}} {{dN(\theta',\varepsilon ,E_{\rm e})}\over{d\varepsilon_\gamma d\varepsilon dt'd\Omega'}}d\varepsilon dE_{\rm e},
\label{eq3}
\end{eqnarray}
\noindent
where $dN(\theta',\varepsilon ,E_{\rm e})/(d\varepsilon_\gamma d\varepsilon dt'd\Omega')$ is
the $\gamma$-ray spectrum produced by the mono-energetic electrons by scattering soft photons with energy 
$\varepsilon$. It is expressed by the convenient formulas obtained by Aharonian \& Atoyan (1981). These spectra depend on the angle, $\theta'$, which is the angle between the directions of the soft photon and the $\gamma$-ray as measured in the blob II reference frame. In our geometrical case this angle is measured from the jet axis (see Fig.~1). In order to produce $\gamma$-ray photon with energy, $\varepsilon_\gamma$, the electron, soft photon, and $\gamma$-ray has to fulfill the following minimum kinetic condition given by, 
$\varepsilon_{\rm min} = 0.5\varepsilon_\gamma m_{\rm e}c^2/[(1 - \cos\theta)E_{\rm e}^2(1 - \varepsilon_\gamma/E_{\rm e})]$ (see Appendix A in Moderski et al.~2005). If $\varepsilon_{\rm min}$ is greater than the lower range of the soft photon spectrum,
$\varepsilon_{\rm L}$, the integration is performed only over the energy range of soft photons above 
$\varepsilon_{\rm min}$. $\varepsilon_{\rm U}$ is the upper energy range of the soft radiation.
We assume that electrons in the blob have the energies in the range $E_{\rm min}$ and $E_{\rm max}$. The lower limit on the energy of the electron, $E_{\rm L}$, which is able to produce $\gamma$-ray with energy, 
$\varepsilon_\gamma$, is also obtained from the above kinetic condition by introducing in the above formula $\varepsilon = \varepsilon_{\rm U}$. This lower limit has to be below the maximum energy of available electrons in the blob, $E_{\rm max}$. If $E_{\rm L}$ is above $E_{\rm min}$ then we integrate the electron spectrum only from $E_{\rm L}$. 

The angle $\theta'$ is related to the observation angle of the jet in the observer's reference frame by the following transformation formula, $\cos\theta' = (\cos\theta - \beta_{\rm II})/(1 - \beta_{\rm II}\cos\theta)$, 
where $\theta$ is the observation angle in the observer's reference frame. The $\gamma$-ray spectra are transformed from the blob II reference frame to the observer's reference frame according to 
$E_\gamma^2 dN/(dE_\gamma dtd\Omega) = D_{\rm II,obs}^4\varepsilon_\gamma^2dN/(d\varepsilon_\gamma dt'd\Omega')$, 
where the Doppler factor is $D_{\rm II,obs} = [\Gamma_{\rm II}(1 - \beta_{\rm II}\cos\theta)]^{-1}$ and 
$E_\gamma = D_{\rm II,obs}\varepsilon_\gamma$. Note that in the case of radio galaxies, such as Cen A and NGC 1275, which are viewed at the range of the angles between $30^\circ-60^\circ$, the Doppler factor is at most of the order of $\sim$2 (for $30^\circ$ and $\Gamma_{\rm II} = 3$) or $\sim$1.15 (for $60^\circ$ and $\Gamma_{\rm II} = 1.12$), see Fig.~2 for the dependence of the Doppler factor on the Lorentz factor of the blob II for selected  values of the observation angles. The radiation, produced within the blob II, is in fact Doppler weakened for the Lorentz factors larger than $\sim$7 in the case of the observation angle $30^\circ$ and $\sim$1.7 for the angle $60^\circ$. Therefore, the emission region in the jet (blob II), responsible for the high energy emission from radio galaxies, which are  viewed at large angles $>30^\circ$, has to move relatively slow.

We assume that electrons in both blobs have the equilibrium spectra well described by the power law function, i.e.
\begin{eqnarray}
dN_{\rm e}/dE_{\rm e} = A E_{\rm e}^{-\beta}. 
\label{eq4}
\end{eqnarray}
\noindent
where $A$ is the normalization coefficient. The spectral index, $\beta$, is linked to the spectral index of the synchrotron radiation produced in the blob I by well known relation
$\beta = 2\alpha - 1$. So then, we assume that the spectral index of electrons does not change for blobs at different location in the jet. In fact, this does not need to be the case. In the general case, different spectral indexes for the electron spectrum in the blob I and blob II can be investigated as well.  

The time scale of the $\gamma$-ray emission from the blob II is determined by the variability time scale of the soft emission produced in the blob I and by the dimension of the blob II. The first time scale is expected to be short since the blob I is moving with the large Lorentz factor, i.e.  
$T_{\rm I}\approx R_{\rm I}/(cD_{\rm I})\approx 100R_{\rm 14.5}/D_{\rm 100}$ s, where the radius of the inner blob is 
$R_{\rm I} = 3\times 10^{14}R_{14.5}$ cm, the Doppler factor of the blob I is $D_{\rm I} = [\Gamma_{\rm I}(1 - \beta_{\rm I})]^{-1} = 100D_{\rm 100}$, and $\Gamma_{\rm I}$ and $\beta_{\rm I}$ are the Lorentz factor and the velocity of the blob I. The radius of the blob is determined by its distance from the base of the jet. For the jet opening angle equal to 0.1 rad, the blob radius is an order of magnitude smaller than its distance from the jet base.
The above estimated time scale variability is of the order of that observed in the BL Lacs such as Mrk~501 and PKS~2155-304. The second time scale depends on the Doppler factor of the blob II, $D_{\rm II,obs}$. Its value, in the case of jets inclined at the angles above $30^\circ$, is not far from unity for mildly relativistic jets. 
The variability time scale of the $\gamma$-ray emission from the blob II is of the order of the light crossing time through the dimension of the blob II, i.e. $T_{\rm II}\sim R_{\rm II}/(cD_{\rm II,obs})$. Therefore, blob II has to have rather small dimension. In the case of NGC 1275 the GeV $\gamma$-ray emission has been observed to vary on the e-folding rise time $8.03\pm 0.24$ hours and on the decay time of $1.21\pm 0.22$ hours (Baghmanyan et al. 2017). This few hours time scale variability corresponds to the dimension of the blob II,
$R_{\rm II}\sim (1.3-8.6)\times 10^{14}D_{\rm II,obs}$~cm. 
This size scale of the emission region is only a few times larger than the Schwarzschild radius of the NGC 1275 SMBH with a mass estimated on $\sim$3.4$\times 10^8$~M$_\odot$ (Wilman et al.~2005). Therefore, either the blob II is located already very close to the jet base, i.e. within a hundred Schwarzschild radii of the SMBH for the jet with the opening angle of the order of degrees, or the GeV $\gamma$-ray emission comes from a different region in the jet, e.g. from the blob I. 
However, the above estimates are becoming less extreme if the new measurements of the black hole mass in NGC 1275, $\sim$3$\times 10^7$~M$_\odot$, are more realistic (Onori et al.~2017).
Another solution of this variability crisis is the assumption of a much smaller inclination angle of the observer to the jet axis than derived in the case of the parsec scale jet in NGC 1275. Note however, that such a very short time scale variability has not been reported during the period of MAGIC observations at sub-TeV $\gamma$-ray energies (and also in the analysis of the simultaneous data from the {\it Fermi}-LAT instrument) when only a hint of variability is observed on a monthly time scale 
(Aleksi\'c et al.~2014). This longer time scale variability corresponds to the size of the stationary emission region of the order of $R\sim 8\times 10^{16}$~cm.

\begin{figure*}
\vskip 5.truecm
\includegraphics{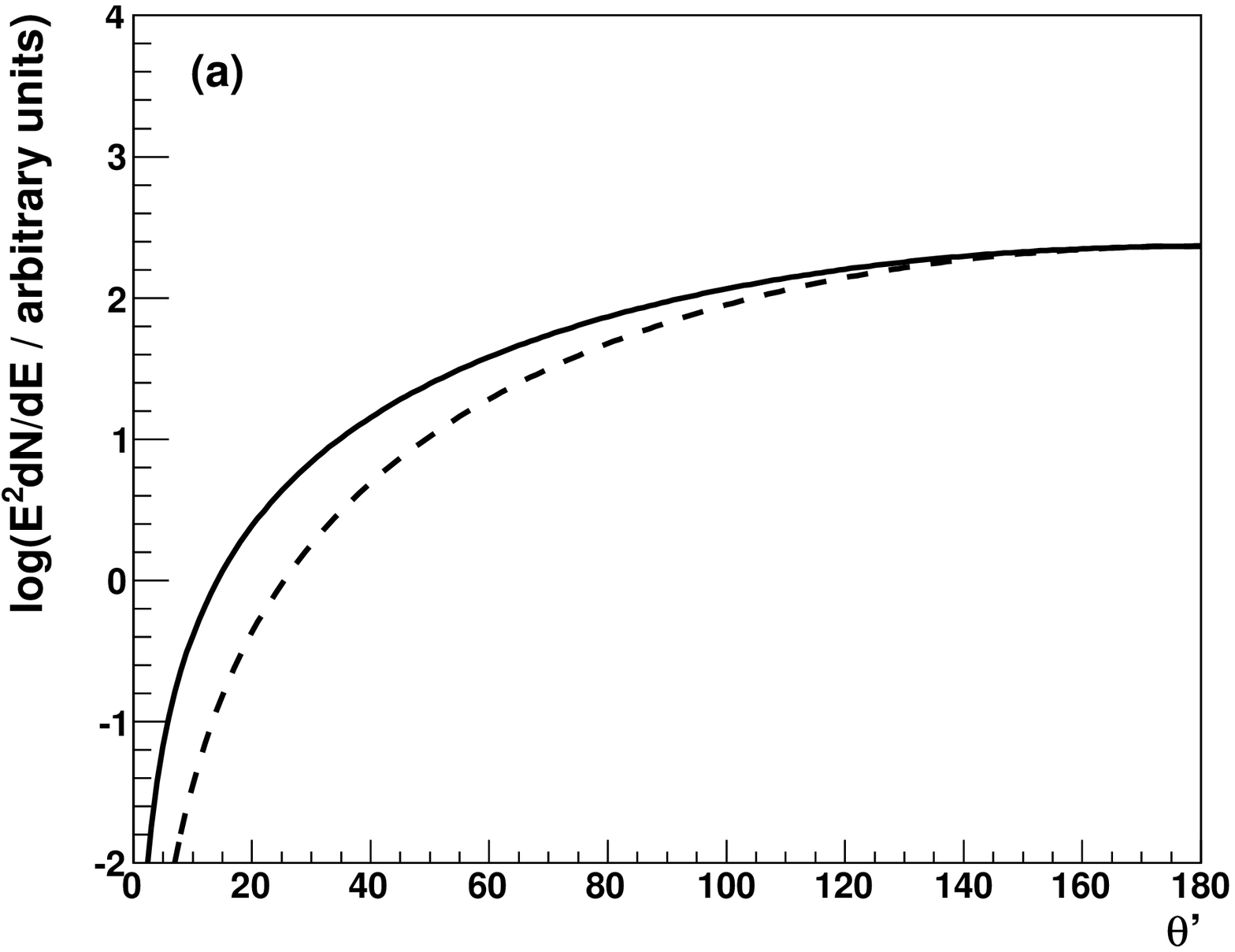}
\includegraphics{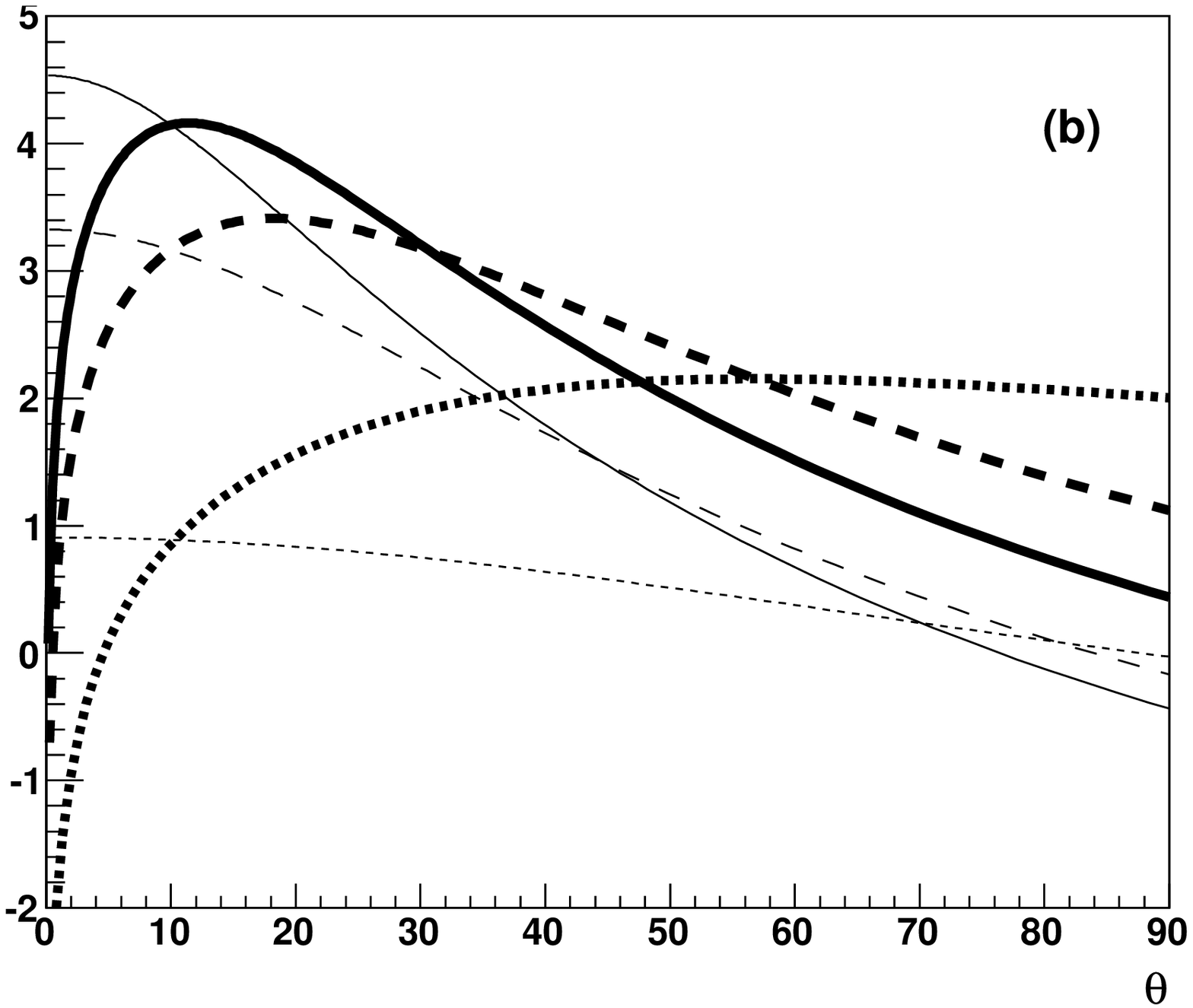}
\includegraphics{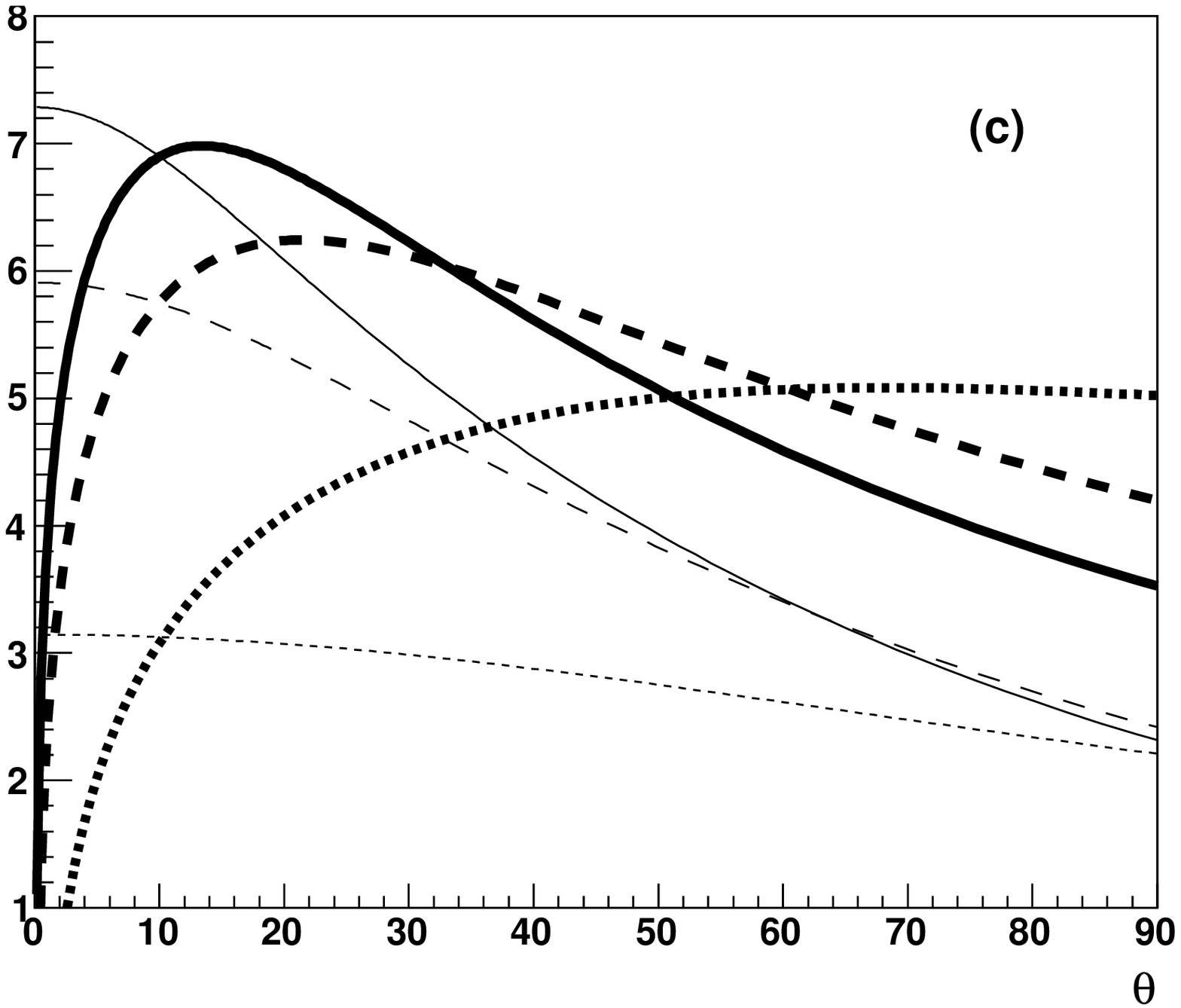}
\caption{The angular distribution of $\gamma$-rays with energy of 1 TeV produced by isotropically distributed electrons, with energy 10 TeV, which comptonize mono-directional soft radiation described by the differential power law spectrum with the spectral index $\alpha$. Soft photons arrive to the electrons in the blob II along the direction described by the angle $\theta' = 180^\circ$ (see Fig.~1). The power in produced $\gamma$-ray photons is shown as a function of the angle 
$\theta'$ in the blob II reference frame for $\alpha = 1.3$ (dashed curve) and 1.8 (solid) (figure a), and 
as a function of the angle measured in the observer's reference frame, $\theta$, for three different values of the blob II Lorentz factor, $\Gamma_{\rm II} = 3$ (thick solid curve), $2$ (thick dashed), and 1.1 (thick dotted) for $\alpha = 1.3$ (figure b) and $\alpha = 1.8$ (c). For comparison, we also show the angular distribution of $\gamma$-ray emission in the case of the isotropic distribution of both soft radiation field and relativistic electrons in the blob II reference frame (see the corresponding thin curves). $\gamma$-ray distributions in both these cases are normalized at the angle $\theta = 10^\circ$.}
\label{fig3}
\end{figure*}
\section{$\gamma$-ray spectra from the blob II}

In the considered scenario, the fast inner blob I produces radiation collimated along the jet axis. The soft energy part of this beamed radiation from the blob I (the synchrotron part)
is comptonized by electrons in the blob II. At first, we investigate the angular dependence of the IC $\gamma$-ray emission from the blob II. It is expected that the essential part of this radiation is emitted at large angles in respect to the jet axis due to the geometry of the IC process occurring in the blob II (the isotropically distributed relativistic electrons and the mono-directional soft radiation arriving from the blob I, see details of the IC scattering process for such a scattering geometry considered by Aharonian \& Atoyan~1981). The spectrum of soft photons is assumed to be of the power law type with the spectral index $\alpha$ and arbitrary normalization. As an example, we show the distribution of $\gamma$-rays with energy of 1 TeV, produced by electrons with the mono-energetic spectrum at energy equal to 10 TeV, in the blob II reference frame (see Fig.~3a). 
These calculations have been performed by integrating Eq.~3 with the electron spectrum given by the Dirac delta function.
As expected, there is a clear deficit of $\gamma$-ray emission in the direction along the jet since soft photons arrive to the blob II exactly from this direction (see Fig.~1). This deficit is larger for the soft radiation with the flatter spectrum, see results for $\alpha = 1.3$ (dashed curve) and 1.8 (solid). After transformation to the observer's reference frame, the $\gamma$-ray emission shows the maximum at a specific angle which depends on the Lorentz factor of the blob II. It is close to $\sim$11$^\circ$ for $\Gamma_{\rm II} = 3$. It moves to $\sim$18$^\circ$ for $\Gamma_{\rm II} = 2$, and to $\sim$57$^\circ$ for $\Gamma_{\rm II} = 1.1$ for $\alpha = 1.3$ (Fig.~3b). These optimal angles are shifted to $13.5^\circ$,  $21.5^\circ$, and $70^\circ$ for $\alpha = 1.8$, respectively (Fig.~3c). Note that the highest level of the 
$\gamma$-ray emission in the range of the observation angles between $30^\circ-60^\circ$, i.e. the range of angles derived for the radio galaxies Cen A and NGC 1275, is observed for the Lorentz factor of the blob II close to $\Gamma_{\rm II} =2$. Therefore, the $\gamma$-ray emission observed in these radio galaxies has to originate in a relatively slowly moving emission regions. For the comparison, we also show the angular distribution of the $\gamma$-ray emission in the observer's reference frame in the case of the isotropic soft radiation field and isotropic electrons in the blob II (see thin curves in Figs.~3b and 3c). These two distributions are normalized at the angle $\theta = 10^\circ$, which is considered as the transition angle for which the AGN is classified either as the BL Lac type or the radio galaxy. Note the clear differences in  the angular distribution of the $\gamma$-ray emission between considered here anisotropic model (thick curves) and the fully isotropic IC model (thin curves). External blob radiation model produces $\gamma$-rays which are clearly more preferentially emitted at large angles to the jet axis.   

\begin{figure*}[t]
\vskip 4.5truecm
\includegraphics{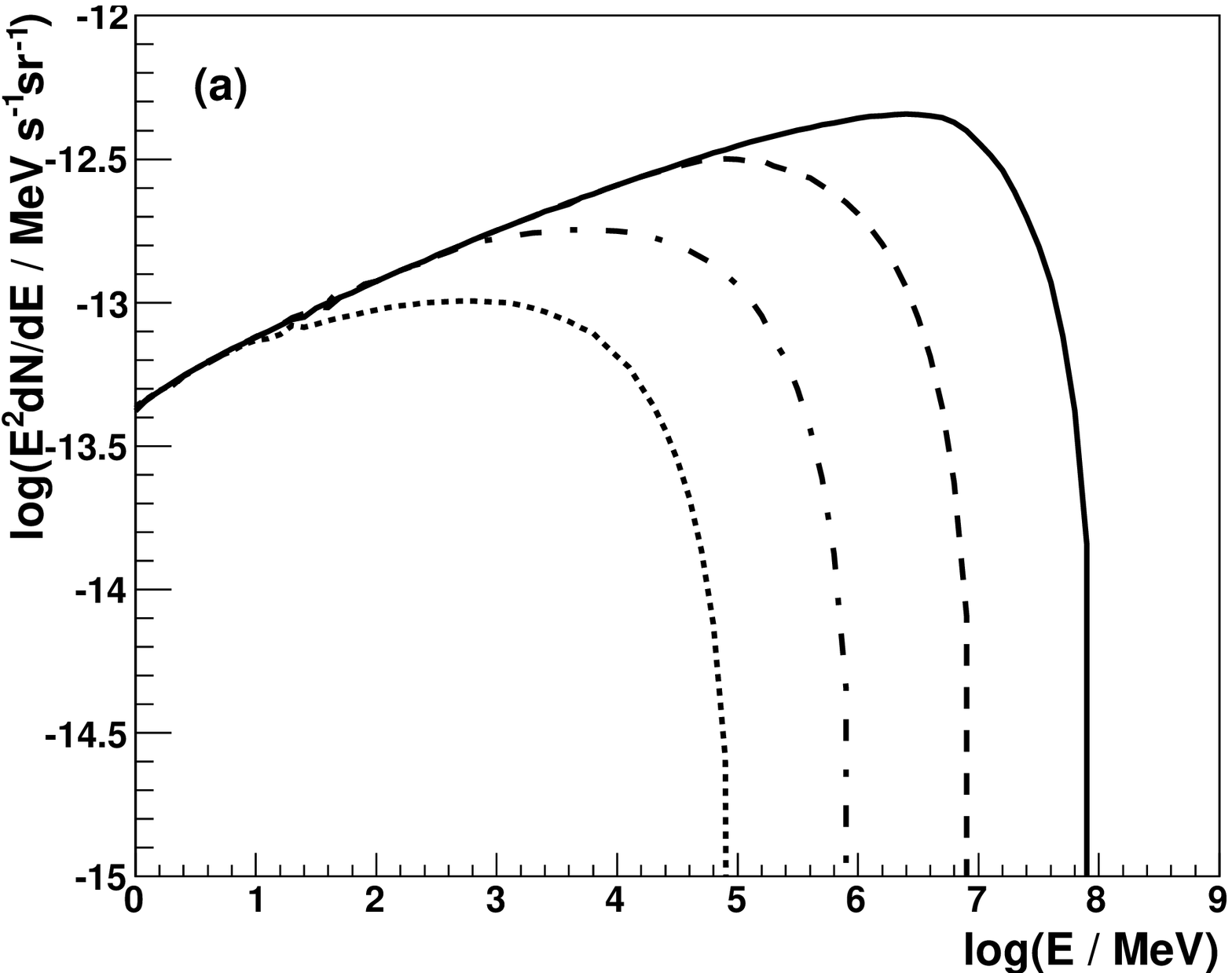}
\includegraphics{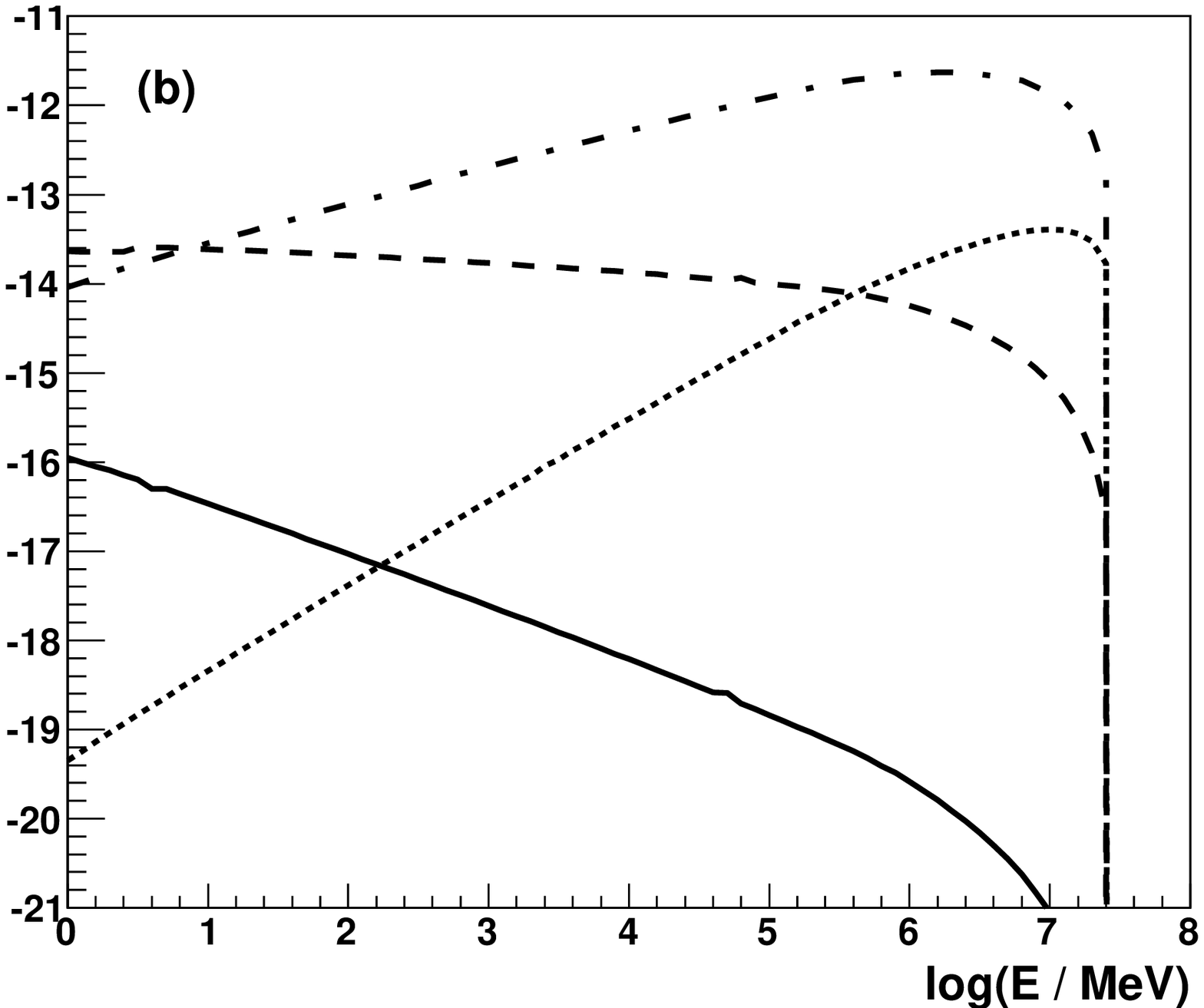}
\includegraphics{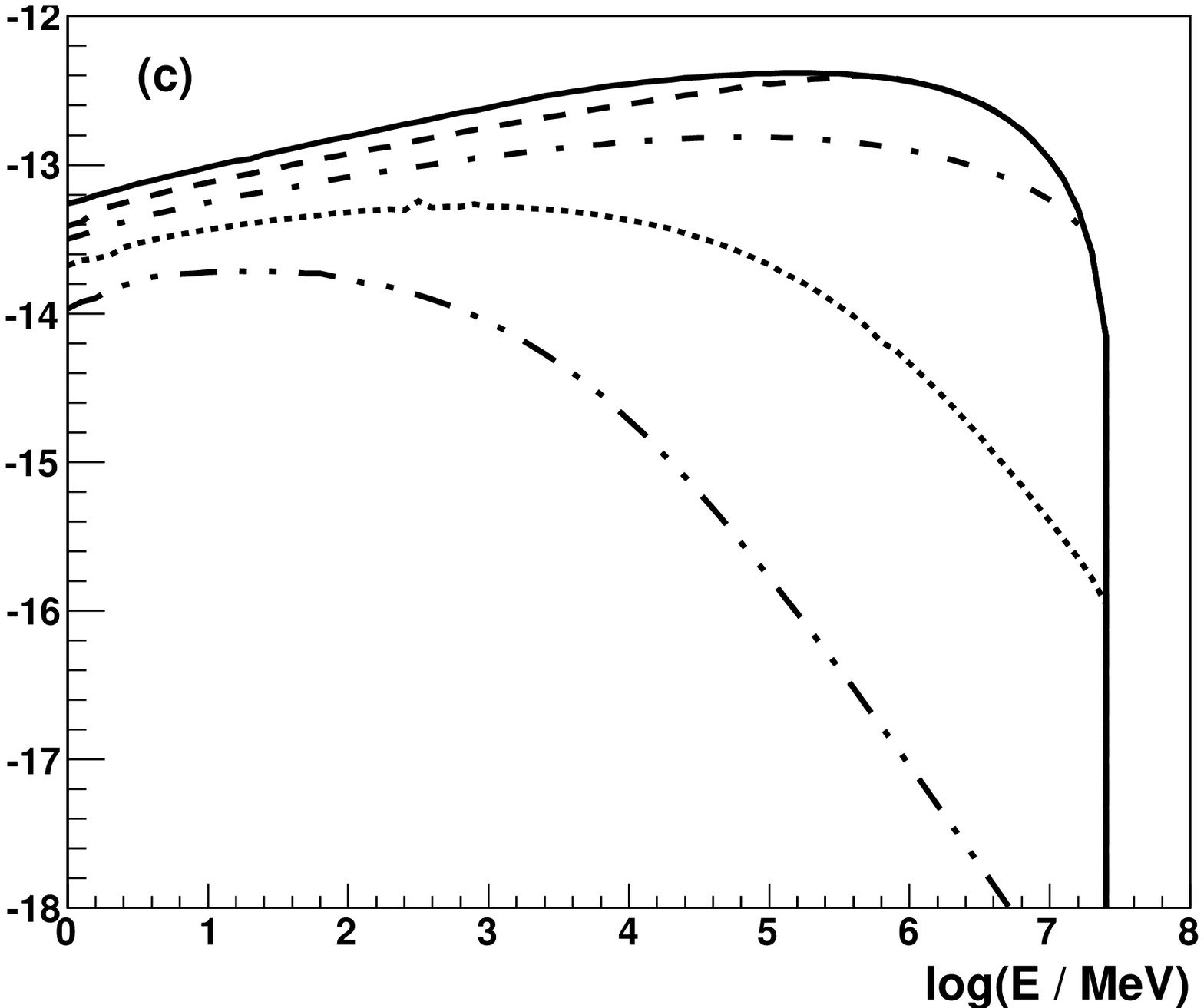}
\includegraphics{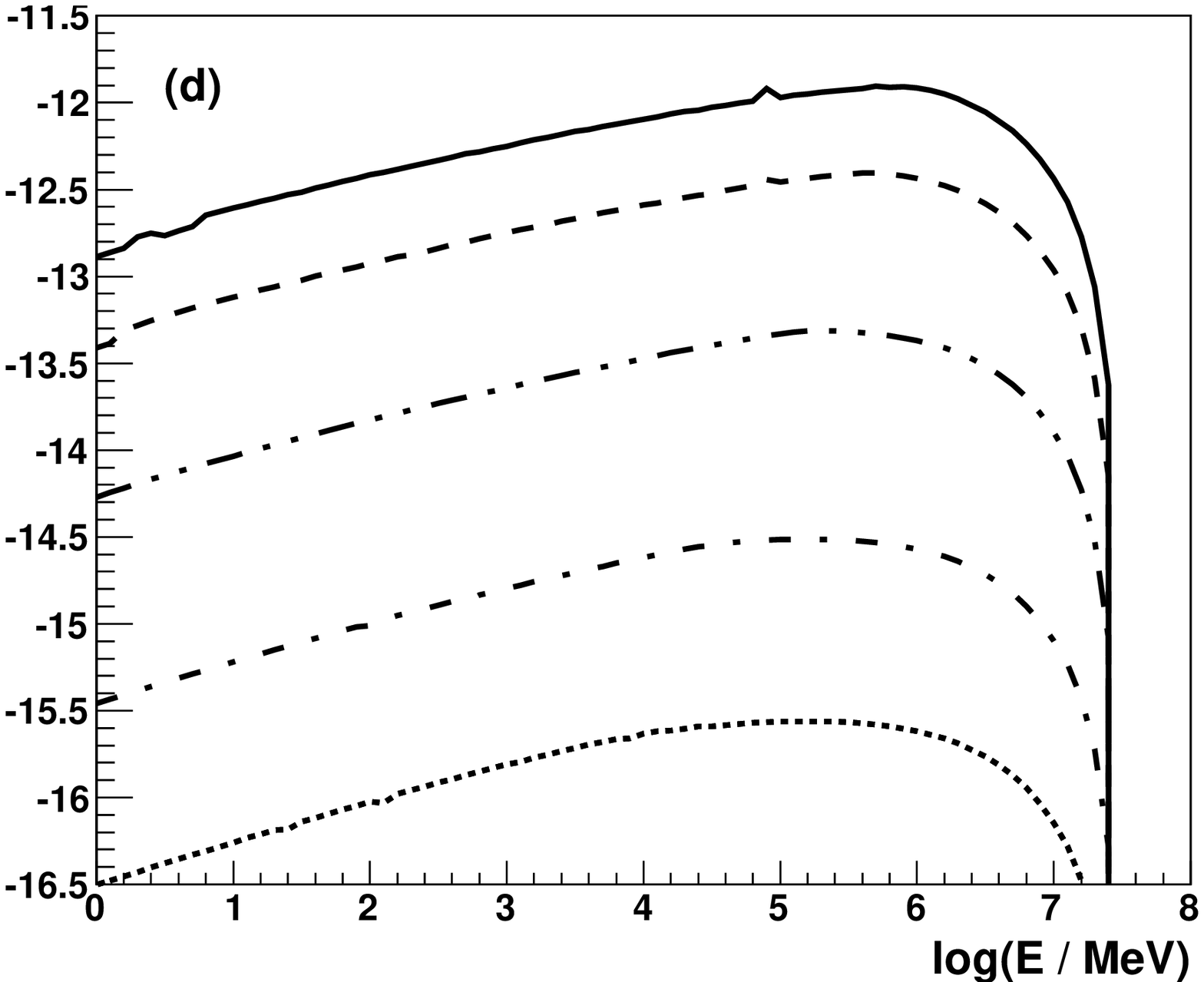}
\caption{SED (Spectral Energy Distribution) of the $\gamma$-rays produced by isotropically distributed electrons in the outer blob II. Electrons IC up-scatter mono-directional soft radiation from the inner blob I. The soft radiation has the simple power law spectrum with spectral index $\alpha$ between $\varepsilon_{\rm min}$ and $\varepsilon_{\rm max}$. Electrons have the power law spectrum, with the spectral index $\beta = 2\alpha - 1$ between $E_{\rm e}^{\rm max}$ and $E_{\rm e}^{\rm min}$, normalized to 1 MeV. We assume that soft photons are produced by electrons in the synchrotron process. Electrons in both blobs are accelerated in a similar way. 
SED as a function of the maximum energy of relativistic electrons, $E_{\rm e}^{\rm max} = 10^8$ MeV (solid), $10^7$ MeV (dashed), $10^6$ MeV (dot-dashed), $10^5$ MeV (dotted) (a), for different spectral indexes of soft photons, $\alpha = 2.5$ (solid), 2.01 (dashed), 1.5 (dot-dashed), and 1. (dotted) (b), for different low energy cut-offs in synchrotron spectrum, 
$\varepsilon_{\rm min} = 10^{-12}$ MeV (solid), $10^{-10}$ MeV (dashed), $10^{-8}$ MeV (dot-dashed), $10^{-6}$ MeV (dotted), $10^{-4}$ MeV (dot-dot-dashed) (c),  and for different emission angles of the $\gamma$-ray photons in the blob II reference frame, $\theta' = 180^\circ$ (solid), $90^\circ$ (dashed), $45^\circ$ (dot-dot-dashed), $20^\circ$ (dot-dashed), and $10^\circ$ (dotted) (d). Unless specified differently, the other parameters of the model are $\alpha = 1.75$, $\theta = 90^0$, $E_{\rm min} = 10$ MeV, $E_{\rm max} = 3\times 10^7$ MeV, $\varepsilon_{\rm min} = 10^{-10}$ MeV and $\varepsilon_{\rm max} = 0.01$ MeV. The spectrum of the mono-directional soft radiation is defined by Eq.~1 and~Eq.~2 with the photon flux at the observer normalized to 1~MeV~cm$^{-3}$.
}
\label{fig4}
\end{figure*}

The $\gamma$-ray spectra produced in the blob II are investigated more systematically in the blob II reference frame on Fig. 4. We show how the $\gamma$-ray spectra depend on different parameters describing this model such as the maximum energies of electrons injected with the power law spectrum (Fig.~4a), spectral index of electrons (Fig.~4b), different low energy cut-offs in the soft synchrotron spectrum arriving to the blob II from the blob I (Fig.~4c), and different emission angles in the blob II reference frame (Fig.~4d). In these calculations, we assume that electrons are accelerated with the power law spectrum. Their spectral index, $\beta$, is related to the spectral index of the soft synchrotron radiation, $\alpha$, according to  
$\beta = 2\alpha -1$. The spectra of soft photons are normalized to 1 MeV~cm$^{-3}$ (Eq.~2) and the spectra of
relativistic electrons to the energy of 1 MeV.  
The dependence of the $\gamma$-ray spectra on the maximum electron energy and on the spectral indexes of soft photons and electrons look quite straightforward. Note that the low energy cut-off in the soft photon spectrum have strong influence on the break in the $\gamma$-ray spectrum (Fig.~4c). If the low energy soft photons are not present, then even energetic electrons cannot produce the high energy $\gamma$-rays due to inefficient scattering of soft photons in the X-ray range. This is caused by the suppression of the cross section in the Klein-Nishina regime. Note also the strong dependence of the level of the spectra on the emission angle of $\gamma$-rays in the blob II reference frame (Fig.~4d). For the considered example parameters, the spectrum emitted at the angle $\theta' = 10^\circ$ is on the level of about four orders of magnitude lower than the spectrum produced at the angle $\theta' = 180^\circ$. 

\begin{figure*}[t]
\vskip 4.5truecm
\includegraphics{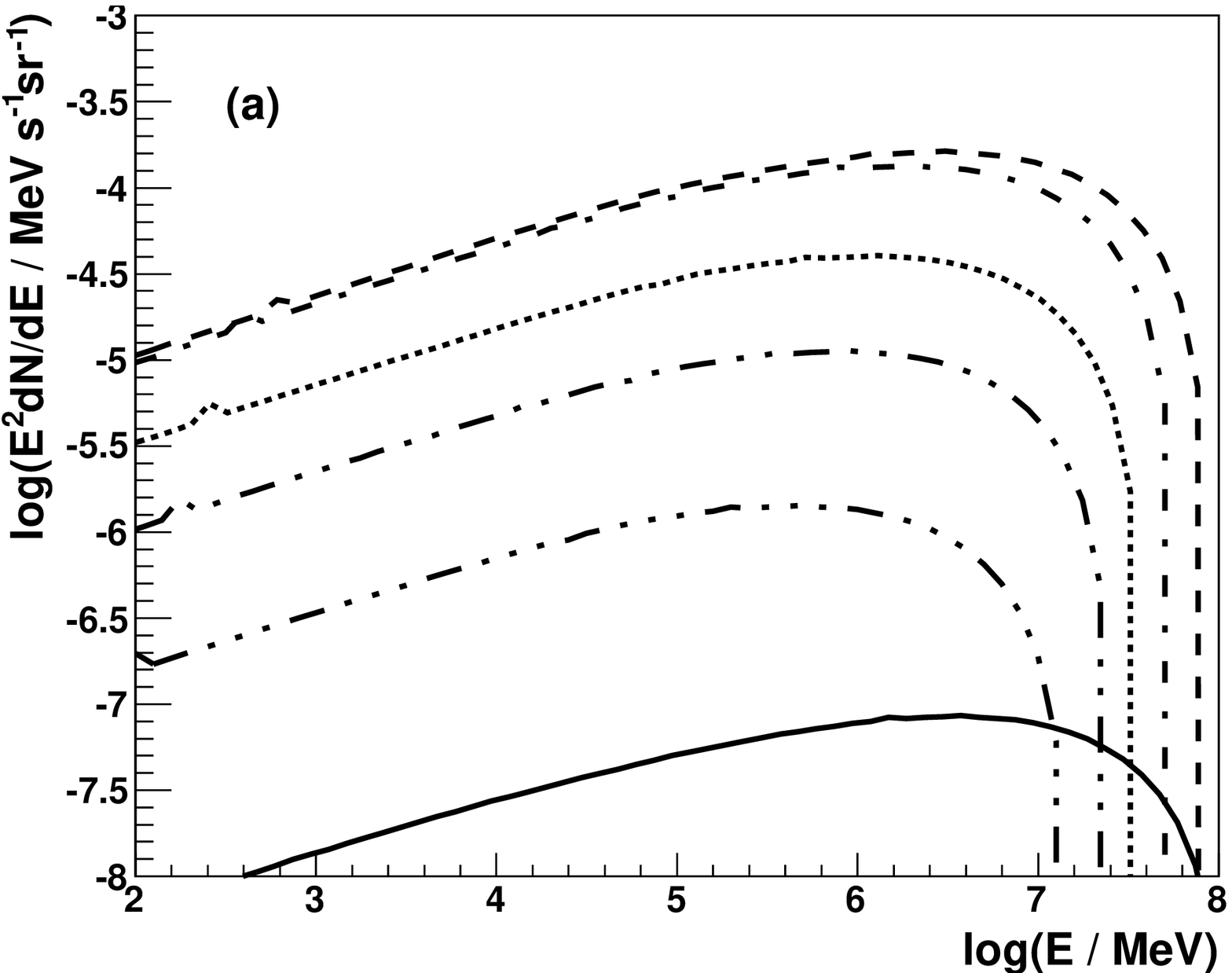}
\includegraphics{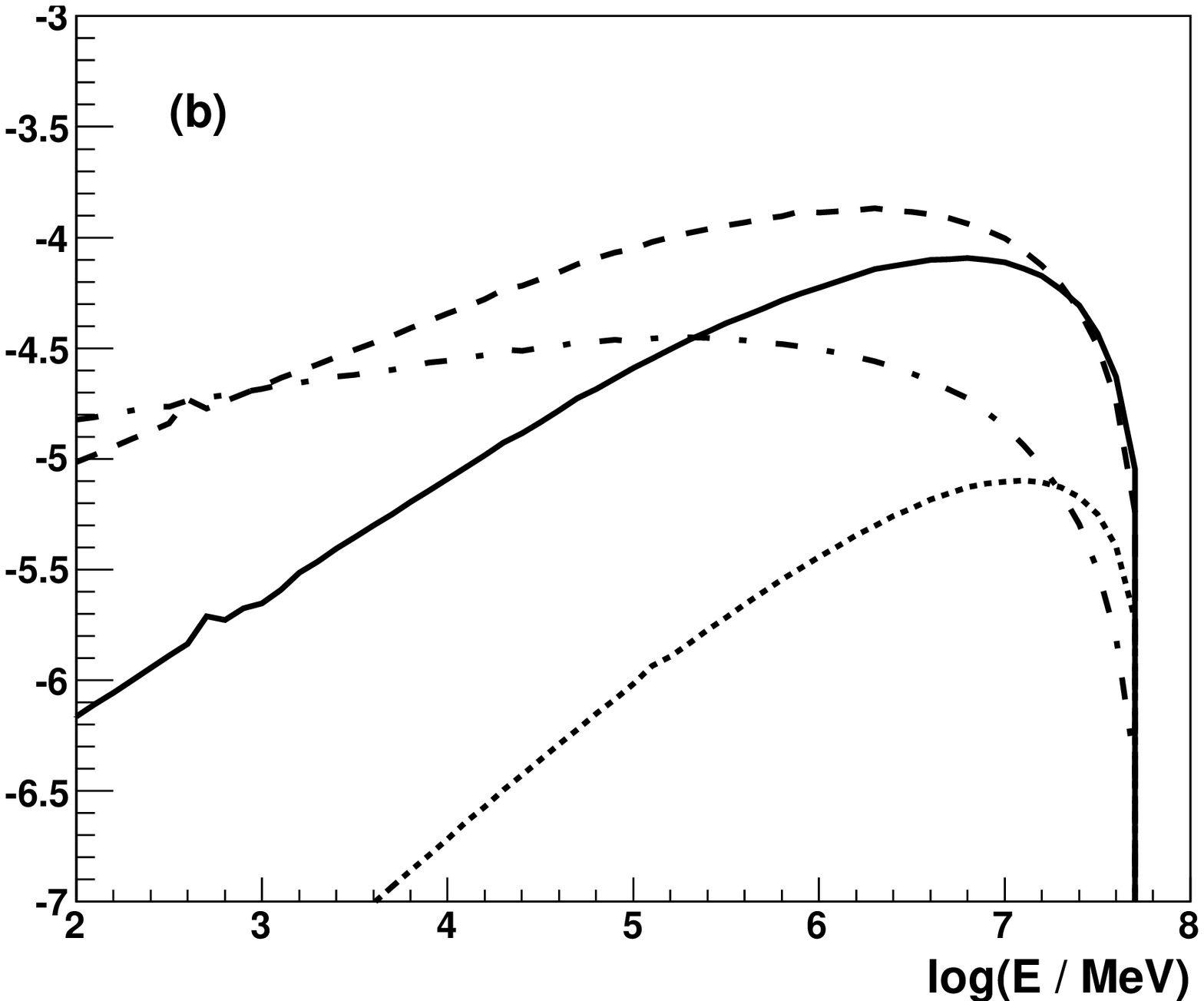}
\includegraphics{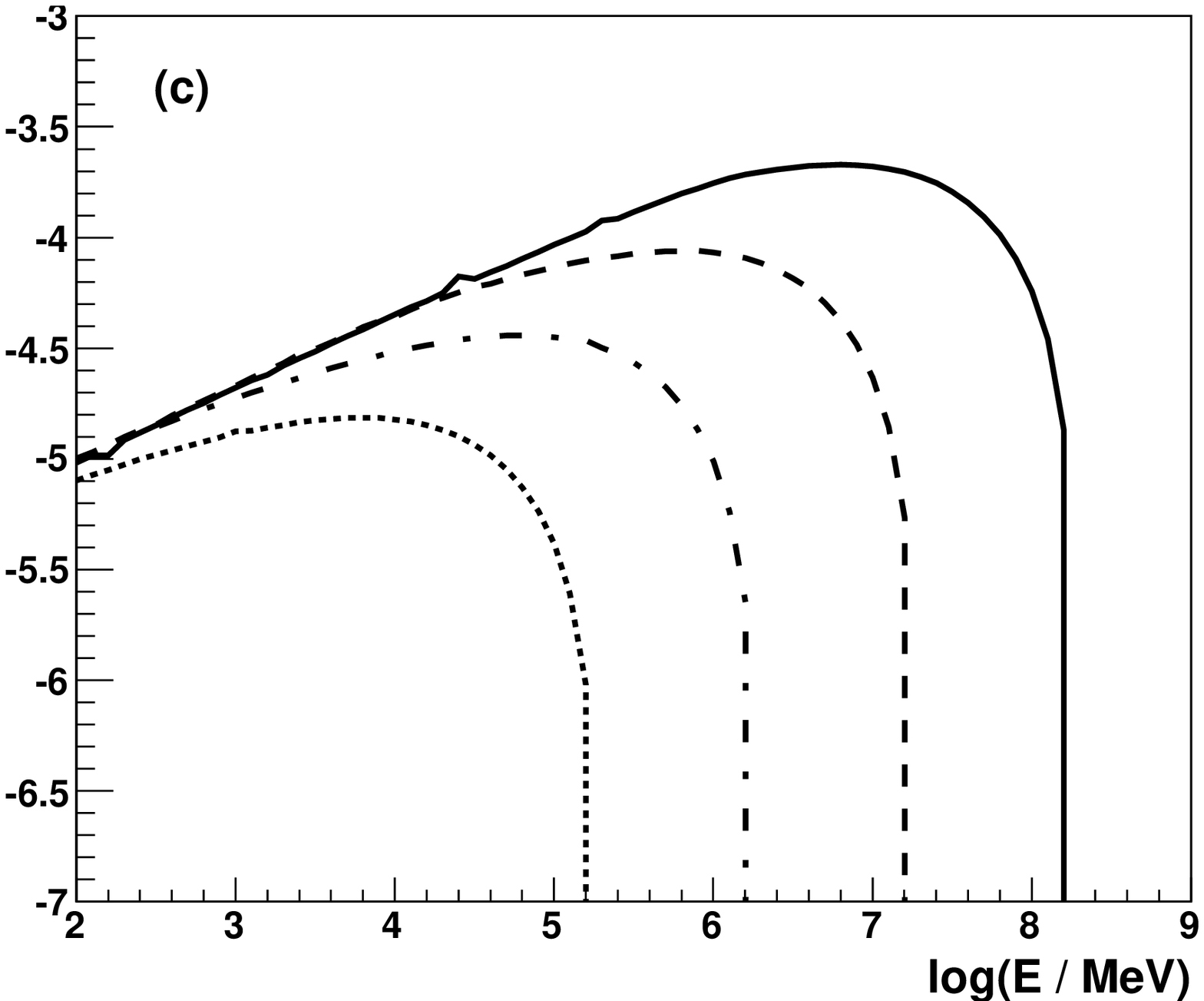}
\includegraphics{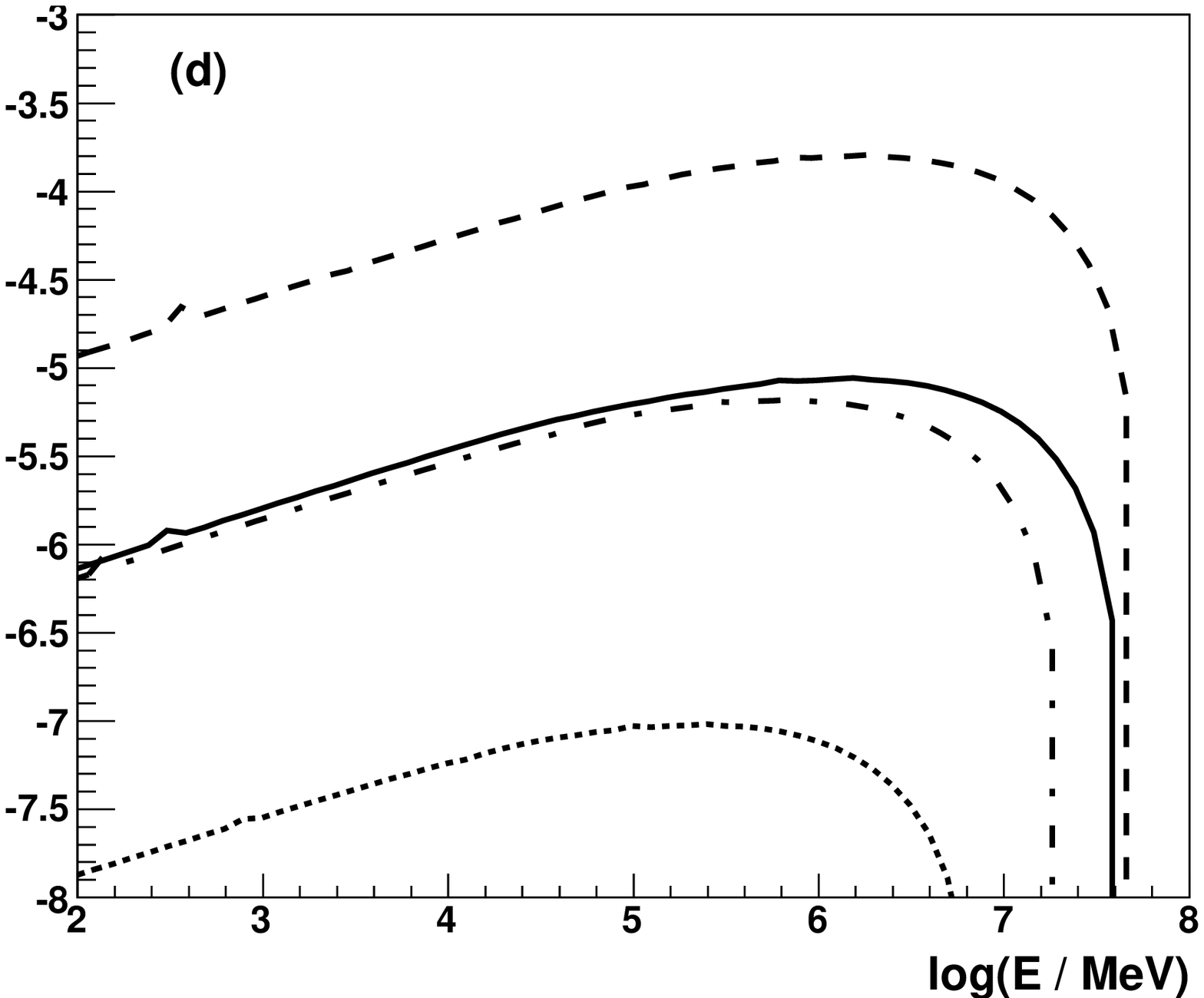}
\caption{SED of $\gamma$-rays for different observation angles in the observer's reference frame (after transforming from the blob II reference frame with the Doppler factor $D_{\rm II,obs} = [\Gamma_{\rm II}(1 - \beta_{\rm II}\cos\theta)]^{-1}$), $\theta = 1^0$ (solid), $15^0$ (dashed), $30^0$ (dot-dashed), $45^0$  (dotted), $60^0$ (dot-dot-dashed), and $90^0$ (dot-dot-dot-dashed) (figure a); for different spectral indexes of the soft radiation field produced in blob I $\alpha = 1.2$ (dotted), 1.4 (solid), 1.6 (dashed), and 1.8 (dot-dashed) (b); for different maximum energies of electrons in the blob II reference frame, $E_{\rm max} = 10^8$ MeV (solid), $10^7$ MeV (dashed), $10^6$ MeV (dot-dashed), and $10^5$ MeV (dotted) (c); and for different Lorentz factors of the blob II equal to $\Gamma_{\rm II} = 1.155$  (solid), 3 (dashed), 10 (dot-dashed), and 30 (dotted) (d). Unless specified differently, the other parameters of the model are $\alpha = 1.6$, $\theta = 30^0$, $E_{\rm max} = 3\times 10^7$ MeV, $\varepsilon_{\rm min} = 10^{-11}$ MeV, $\varepsilon_{\rm max} = 0.2/D_{\rm II,obs}$ MeV, and $\Gamma_{\rm II} = 2$. Electrons are injected isotropically in the blob II reference frame with the power law spectrum, spectral index $\beta = 2\alpha - 1$, which is normalized to 1 MeV above electron energy equal to 10 MeV. The spectrum of the mono-directional soft radiation is defined by Eq.~1 and~2 and normalized to the isotropic equivalent power in the observer's reference frame equal to $10^{45}$ erg~s$^{-1}$ assuming the location of the blob II at the distance of $z = 10^{16}$ cm (see Fig.~1).}
\label{fig5}
\end{figure*}

In order to perform a more realistic calculations of the $\gamma$-ray spectra expected in such a model,
we fix the spectrum of the soft emission from blob I. It is assumed to be similar to that observed in classical BL Lacs. As an example, we apply the level of the synchrotron spectrum observed during one of the flares in Mrk 501, i.e. its isotropic equivalent soft radiation luminosity is $10^{45}$~erg~s$^{-1}$ and the spectrum is described by Eq.~1 and 2. The spectral index of this emission is approximated by $\alpha = 1.6$. The spectral index, $\beta$,  of the power law spectrum of isotropic electrons is assumed to be linked to the spectrum of soft photons as expected in the synchrotron process, i.e. $\beta = 2\alpha -1$. The example calculations of the 
$\gamma$-ray spectra produced in the blob II are shown in Fig.~5. 
In general the SED of produced $\gamma$-rays is shaped by the three competing effects. At first, the soft radiation field produced in the blob I, and detected in case of the observer located on the jet axis (the BL Lac type emission), is reduced by the Doppler factor $D_{\rm II} = [\Gamma_{\rm II}(1 +\beta_{\rm II})]^{-1}$ since this mono-directional radiation approaches the blob II from behind, i.e. from the base of the jet. At second, the up-scattering of the mono-directional soft photons by the isotropic relativistic electrons in the blob II reference frame occurs strongly anisotropically, i.e. preferentially in the direction of up-coming soft photons consistent with the direction of propagation of the blob I. This effect is responsible for the emission of $\gamma$-rays preferentially in the opposite direction to the direction of motion of the blob II.
At third, the $\gamma$-rays from the blob II are Doppler boosted with the Doppler factor $D_{\rm II,obs} = [\Gamma_{\rm II}(1 - \beta_{\rm II}\cos\theta)]^{-1}$. This effect can either enhance the $\gamma$-ray fluxes in the observer's reference frame or significantly reduce them, depending on the value of the observation angle, 
$\theta$, in the observer's reference frame. The combination of these three effects is responsible for the features of the $\gamma$-ray SED shown in Fig.~5. 

Note the interesting dependence of the spectra on the observation angle, $\theta$, in the observer's reference frame. In contrast to the homogeneous SSC model, the $\gamma$-ray spectra, produced in terms of our model, do not show maximum fluxes for the angle $\theta = 0^\circ$ but for the specific range of angles which is characteristic for the radio type galaxies (see spectra for the range of angles $\theta = 15^\circ - 30^\circ$ in Fig.~5a). Therefore, we conclude that such a model can explain more naturally the appearance of the high energy $\gamma$-ray emission from the radio galaxies which are viewed at relatively large angles.     
The $\gamma$-ray spectra produced in our model do not behave systematically also on the change of the Lorentz factor of the blob II (see Fig.~5d). The largest $\gamma$-ray fluxes are expected for the intermediate values of the Lorentz factors of the emission region (blob II) of the order of $\Gamma_{\rm II} = 2-3$. Therefore, the optimal conditions for the $\gamma$-ray emission at large angles are provided by the blobs moving with the intermediate Lorentz factors. Then, the Doppler factors, characterizing such blobs, obtain the optimal values as already discussed above (see Fig.~2). 
On the other hand, dependence of the $\gamma$-ray SED on other parameters is more systematic. For example, the spectra show simple dependence on the spectral indexes of the soft radiation and the relativistic electrons (see Fig.~5b) and on the maximum energy in the spectrum of relativistic electrons (see Fig.~5c).

\section{Interpretation of NGC 1275}

\begin{figure*}
\vskip 5.truecm
\includegraphics{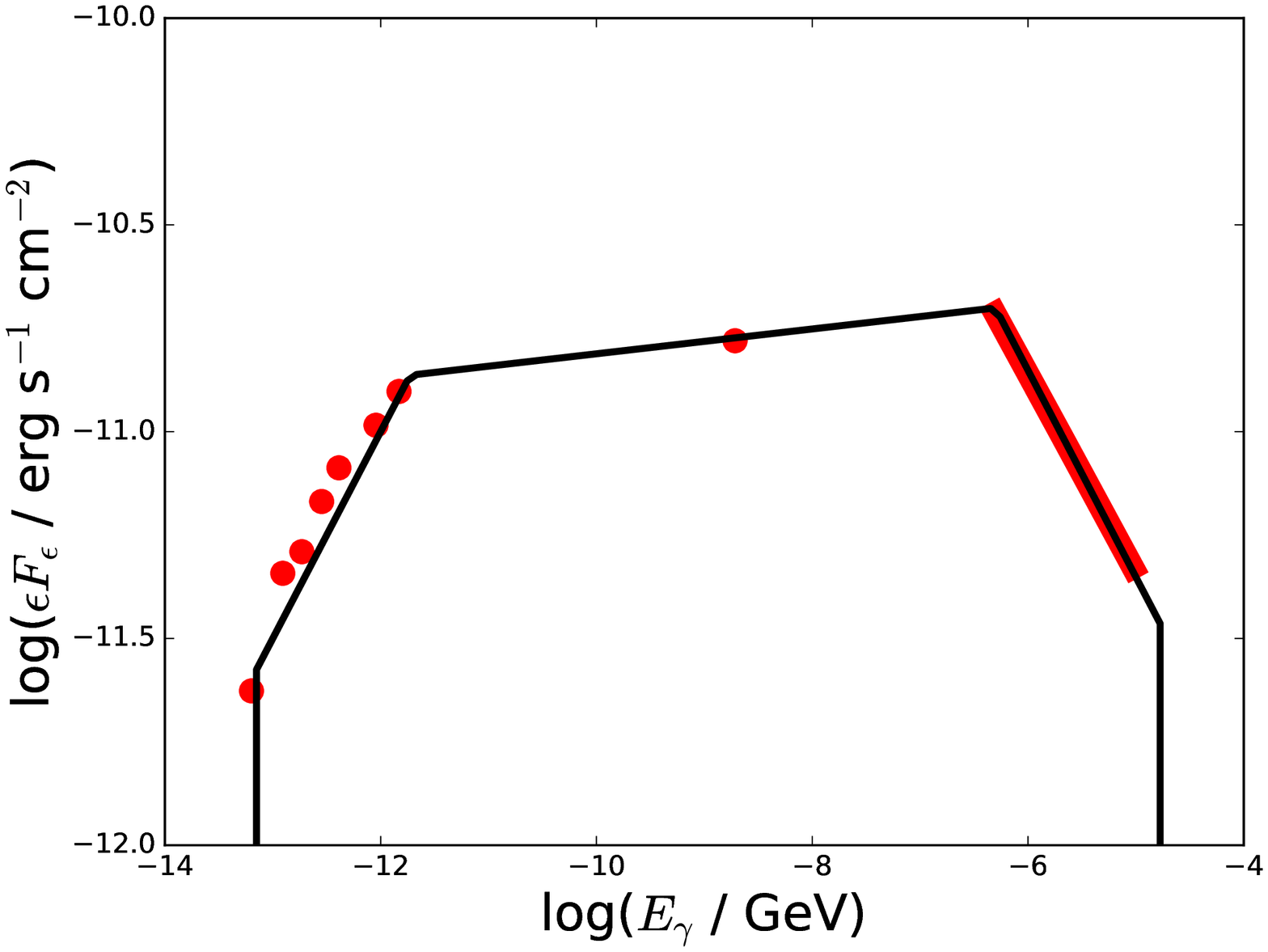}
\includegraphics{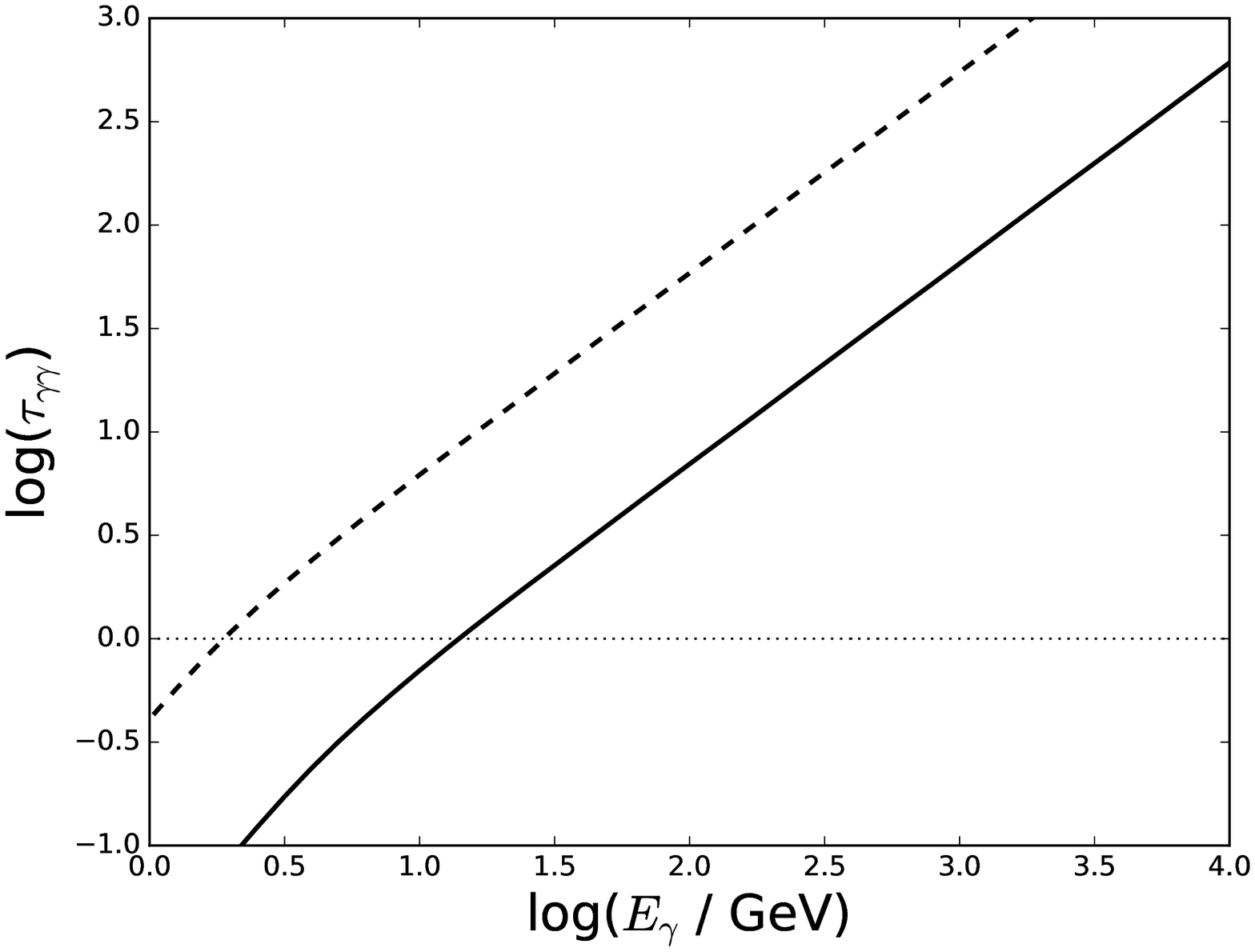}
\caption{The soft radiation field within the blob in the jet of NGC 1275 is constrained in the optical (KVA) and X-ray energies ({\it Chandra}) which are quasi-simultaneous with the MAGIC and {\it Fermi}-LAT observations (marked by red colors in figure on the left, data interpolated from Fig.~10 in Aleksi\'c et al.~2014). Its synchrotron spectrum is characterized by the broken power law marked by the black curve. The optical depths for $\gamma$-ray photons in the soft radiation, observed in the low energy bump
of the spectrum from NGC 1275, are shown as a function of $\gamma$-ray energy (on the right). 
The blob II has the radius $R_{\rm II} = 3\times 10^{15}$ cm and the Doppler factor equal to $D_{\rm II,obs} = 2$ (solid curve) and 1.3 (dashed curve).}
\label{fig6}
\end{figure*}

NGC 1275 is the radio galaxy in the Perseus cluster which hosts a super-massive black hole with the mass of 
$\sim (3.4-8)\times 10^8$~M$_\odot$ (Wilman et al.~2005, Scharw\"achter et al. 2013).
The observed isotropic equivalent  luminosity in the soft radiation observed from NGC 1275 is of the order of $4\pi d_{\rm L}^2F(\varepsilon) = 1.5\times 10^{44}$ erg~s$^{-1}$, where the luminosity distance to NGC 1275 is assumed to be 70 Mpc for the redshift $z = 0.017559$ (see NED: http://ned.ipac.caltech.edu/ and Fig.~6 on the left).
However this energy is emitted at a relatively large angle to the jet axis, i.e. $\theta = 30^\circ - 50^\circ$.
Due to the beaming effects, the synchrotron emission from the inner blob I is expected to be larger in the direction along the jet axis, i.e. comparable to that observed in BL Lac type AGNs observed along the jet. According to our model this soft photon emission along the jet axis serves as a target for the relativistic electrons in the outer blob II. 
As an example, we perform the calculations of the $\gamma$-ray emission from the blob II assuming that the isotropic equivalent power in the soft radiation field, as observed along the jet axis, is in the range $3\times 10^{45} - 10^{46}$ erg~s$^{-1}$.
For this power, we obtain the density of the soft photons in the blob II by using Eq.~2 and assuming that the blob II is at the distance of $3\times 10^{16}$ cm, or $3\times 10^{17}$ cm, from the base of the jet. The spectral index of this soft radiation has been fixed on $\alpha = 2$ between $\varepsilon_{\rm min} = 10^{-8}$ MeV and $\varepsilon_{\rm max} = 10^{-5}$ MeV in order to be consistent with the observations of NGC 1275 (see NED: http://ned.ipac.caltech.edu/).
Note that the soft synchrotron spectrum in NGC 1275 extends also outside the energy range determined by 
$\varepsilon_{\rm min}$ and $\varepsilon_{\rm max}$ (although with different spectral indexes).
We have checked that the scattering of the soft radiation field outside the energy range $\varepsilon_{\rm min}$ and $\varepsilon_{\rm max}$ by relativistic electrons has only a minor effect on the $\gamma$-ray spectrum. 
We perform the example calculations of the $\gamma$-ray spectra assuming two limiting values for the observation angles of the jet in NGC 1275, i.e. $\theta = 30^\circ$ and $50^\circ$. 

These $\gamma$-rays have to propagate through the synchrotron radiation field produced by electrons 
within the blob II.
In order to find out whether their absorption is important, we calculate the optical depths for the $\gamma$-rays in the radiation field of the homogeneous blob II. Unfortunately, the low energy bump in the NGC 1275 spectrum is poorly constrained by the observations. We apply the Planck data, the optical KVA and Chandra observations which are nearly simultaneous with the MAGIC and Fermi-LAT observations as reported in Fig.~10 in Aleksi\'c et al.~(2014).
The optical depths are calculated following the procedure and the standard formula shown in Sect.~2.2 in Bednarek \& Protheroe~(1999), see also Dondi \& Ghisellini~(1995). 
They are shown in Fig.~6 for the radius of the blob II equal to $3\times 10^{15}$ cm, and two values of the the Doppler factor of the blob II equal to $D_{\rm II, obs} = 2$ and 1.3.
We also show the soft radiation field produced within the blob II for which these optical depths have been obtained.
It is clear that the absorption effects in the soft radiation field of the blob II has to be taken into account for the case of the blob II with the radius equal to $3\times 10^{15}$ cm. 
Note that the optical depth is inversely proportional to the radius of the blob. Therefore, for the blob II with the radius $3\times 10^{16}$ cm, the absorption effects becomes much smaller. 
We include the absorption effects on the $\gamma$-ray spectra, produced in the homogeneous blob II, by introducing the reduction factor equal to $R(E_\gamma) = \{1 - \exp[-\tau(E_\gamma)]\}/\tau(E_\gamma)$.

In fact, produced $\gamma$-rays might be also absorbed in the soft radiation field from the blob I. It is expected that this absorption plays minor role in the absorption process of $\gamma$-rays since, 
due to the geometry of this radiation, i.e. mono-directional distribution approaching from the direction behind the blob II. The threshold condition for the $\gamma-\gamma$ absorption is clearly larger, i.e. 
$\varepsilon > 2m_e^2c^4/E_\gamma /(1 - \cos\theta_{\rm obs})$, where $m_{\rm e}c^2$ is the electron rest energy. For the observation angle $\theta_{\rm obs} = 30^\circ$, the threshold is ~7.5 larger that 
results in the reduced density of soft photons from the blob I by a factor of $\sim (7.5)^2\sim 55$, assuming the soft differential power law spectrum with the index -2. Correspondingly, for the angle 
$\theta_{\rm obs} = 50^\circ$, this factor becomes $\sim (2.8)^2\sim 7.8$. 
Therefore, we conclude that the absorption effects of $\gamma$-rays in the soft radiation from the blob I 
should be usually less important than the absorption of $\gamma$-rays in the soft radiation produced within the blob II.

\begin{table}
  \caption{The total energy in relativistic electrons and the magnetic field strength within the large and small blob II}
  \begin{tabular}{lll} 
\hline 
\hline 
model & large blob II  &  small blob II   \\
R & $3\times 10^{16}$~cm &  $3\times 10^{15}$~cm  \\
z & $3\times 10^{17}$~cm &  $3\times 10^{16}$~cm  \\
$L_\varepsilon$ & $10^{46}$~erg~s$^{-1}$  &  $3\times 10^{45}$~erg~s$^{-1}$  \\
\hline
$\theta = 30^\circ$ &  $1.2\times 10^{49}$~erg  &  $4\times 10^{47}$~erg \\
B & 5 G & 27 G \\
\hline
$\theta = 35^\circ$ &    &  $10^{48}$~erg \\
B &  & 17 G \\
\hline
$\theta = 50^\circ$ &  $4.7\times 10^{49}$~erg  &   \\
B &  2.5 G &  \\
\hline 
\hline 
\end{tabular}
  \label{tab1}
\end{table}

The $\gamma$-ray spectra, produced in the case of small ($R = 3\times 10^{15}$ cm and $z = 3\times 10^{16}$ cm) and large ($R = 3\times 10^{16}$ cm and $z = 3\times 10^{17}$ cm) blob II, are shown in Fig.~7. Good description of the NGC 1275 $\gamma$-ray spectrum is obtained in the case of a small blob II for the observation angles closer to $\theta\sim 30^\circ$ provided that the total energy in the relativistic electrons in the blob II is of the order of a few $10^{47}$ ergs (see Fig.~6 on the left and Table~1). 
This total energy is obtained from the formula
\begin{eqnarray} 
U_{\rm e} = \int_{E_{\rm min}}^{E_{\rm max}} {{dN_{\rm e}}\over{dE_{\rm e}}}E_{\rm e}dE_{\rm e},
\label{eq5}
\end{eqnarray}
\noindent
where the normalization constant $A$ obtained from the comparison of the observed $\gamma$-ray spectrum from NGC 1275 with the spectrum calculated in terms of our model for the normalization of the constant $A$ equal to unity, and 
$E_{\rm min}$ and $E_{\rm max}$ are the applied minimum and maximum energies of relativistic electrons. 
The electron spectrum is assumed to be well described by a single power law up to TeV energies. The change in the observed sub-TeV $\gamma$-ray spectrum is due to the absorption in the soft radiation from the Blob II. 
The spectral index of the electron spectrum is related to the spectral index of the soft synchrotron radiation 
$\beta = 2\alpha - 1 = 3$. However, the absorption effects in the $\gamma$-ray spectrum calculated for 
the small blob II and
the observation angle $\theta = 50^\circ$ are too strong in order to describe the $\gamma$-ray spectrum from NGC 1275 (Fig.~7). In the case of the large blob II, located at larger distance from the base of the jet, 
the absorption effects are too low to explain the change of the spectral behavior in the GeV-TeV energy range.
Therefore, we consider the scenario in which the spectral change in the $\gamma$-ray spectrum is due to the acceleration of electrons to only sub-TeV energies. With such an assumption, the $\gamma$-ray spectrum from NGC 1275 can be well explained for the range of the observation angles, $\theta = 30^\circ-50^\circ$ (see Fig.~7 on the left). However the requirement on the energy in relativistic electrons, i.e. a few $10^{49}$ erg, is more restrictive (see Table~1). However, even such large total energy in electrons in the blob II
seems to be acceptable keeping in mind that NGC 1275 emits $\sim 10^{44}$ erg~s$^{-1}$ for the dynamical time 
scale of the emission in the jet of the order of $\sim z/c\sim 10^7$ s.
Note that estimated total energy in relativistic electrons in the blob II are anti-correlated with the assumed power in the soft synchrotron radiation and with the square distance of the blob II from the base of the jet. We conclude that in the case of the large blob II the GeV-TeV $\gamma$-ray emission from the radio galaxy NGC 1275 can be well described by the external blob radiation model even if the jet is viewed by the observer at a relatively large angle to the jet axis.

We also calculate the magnetic field strength in terms of the considered models which is required 
in order to produce observable flux of the synchrotron bump. We apply the relativistic electrons
with the total energy (and spectrum) reported in Table 1. The obtained values of the 
magnetic field strength are also reported in Table 1. For the considered models, the magnetic field strength
should be in the range between $B = 2.5$ G to 27 G. 
We conclude that in the case of the large blob II model, viewed at the angle of $50^\circ$,
the energy density of the magnetic field is close to the equipartition to the energy density
of relativistic electrons. In the case of other models, the energy density of the magnetic field 
dominates over the energy density of relativistic electrons. Therefore, magnetic energy can serve
as a source of energy for the acceleration of electrons. Moreover, it is expected that the electrons
are confined within the blob by the magnetic field.

The electrons in the blob II should also produce $\gamma$-rays by componizing soft radiation
produced in the blob II, i.e.  in the synchrotron self-Compton (SSC) process. 
We calculate the $\gamma$-ray spectra produced in the blob II in terms of the SSC model for these same parameters 
as considered in the case of the external blob radiation models with the small and large blob II. The example spectra are shown 
in Fig.~7 for the small blob II observed at the angle $\theta = 35^\circ$ (dot-dashed curve) and for the large blob II at the angles 
$\theta = 30^\circ$ (dot-dashed) and $\theta = 50^\circ$ (dot-dot-dashed). Our calculations show that the $\gamma$-ray spectra produced in terms of the external blob model clearly dominate over the $\gamma$-ray spectra produced within the blob II in terms of the SSC model. This effect is due to the fact that the density of photons from the blob I within the blob II, calculated for the luminosity 
$3\times 10^{45}$ erg~s$^{-1}$ and $10^{46}$ erg~s$^{-1}$, is much stronger than the density of photons produced within the blob II (the observed luminosity of the NGC 1275 $\sim 1.5\times 10^{44}$~erg~s$^{-1}$).

\section{Conclusion}

\begin{figure*}[t]
\vskip 5.truecm
\includegraphics{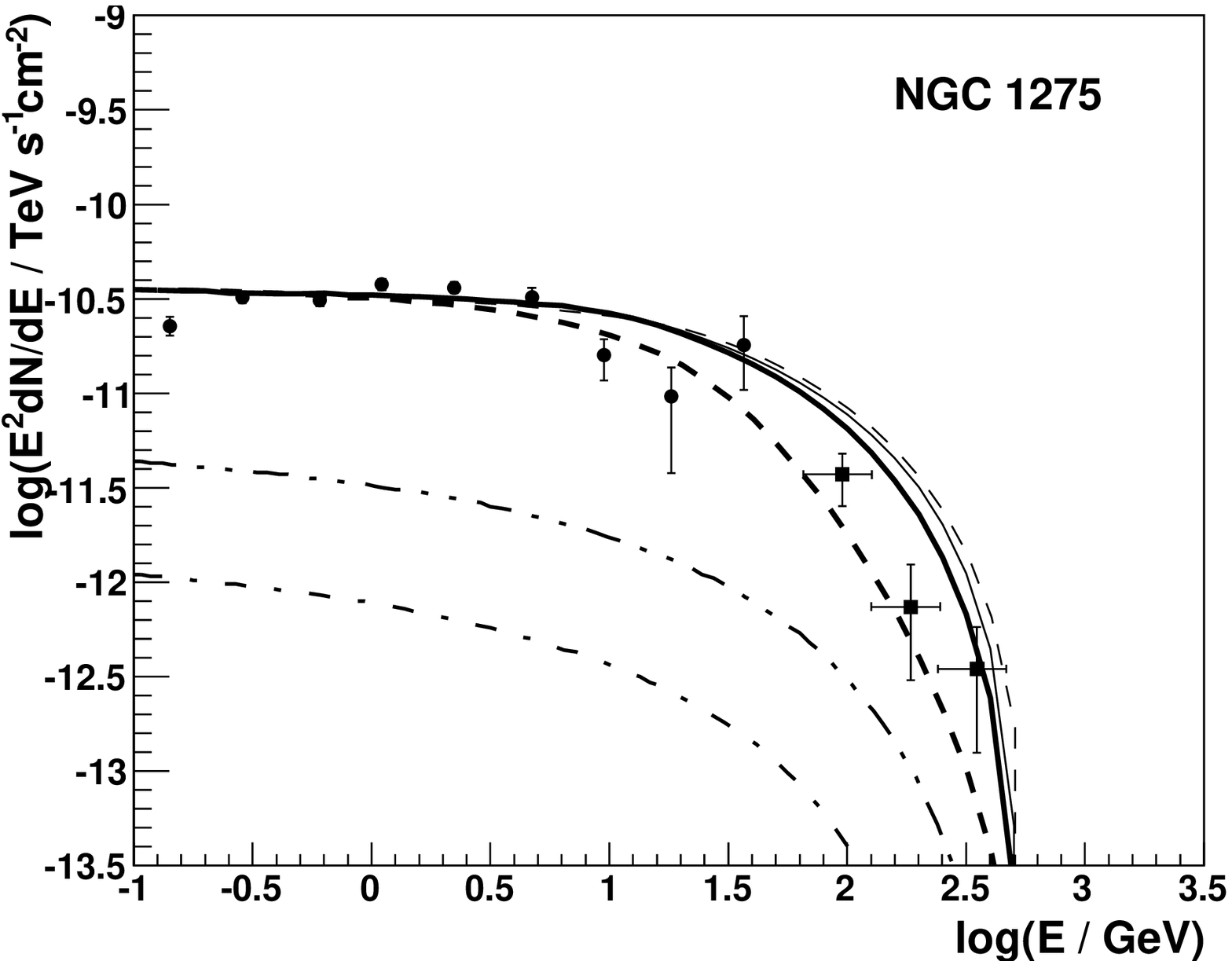}
\includegraphics{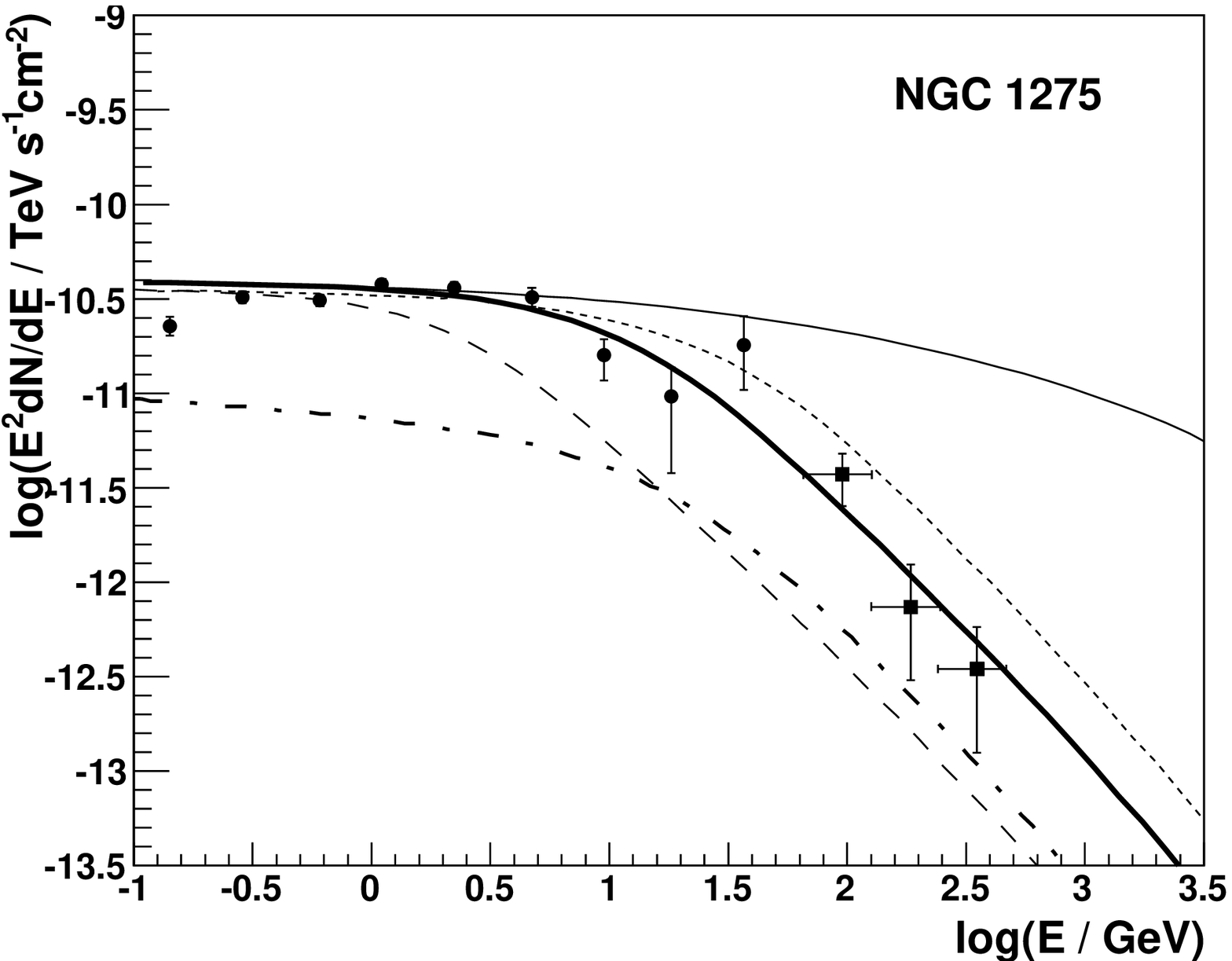}
\caption{Interpretation of the $\gamma$-ray SED observed from the radio galaxy NGC 1275 by the $\gamma$-ray spectra expected in terms of the small (right figure) and large (left figure) external blob radiation model. In the small blob model, the SED is described by the spectra calculated for the angle $\theta = 35^\circ$, the Lorentz factor of the blob I $\Gamma_{\rm II} = 1.7$, with the absorption effects included as described in the main text (thick solid curve). The spectrum before absorption in the blob II radiation is shown by the thin solid curve.  
The spectra obtained for the angles $\theta = 30^\circ$ (and $\Gamma_{\rm II} = 2$) and  $50^\circ$ (and $\Gamma_{\rm II} = 1.5$) are shown by the thin dotted and  dashed curves, respectively. 
In the large blob II model, the SED is well described by the $\gamma$-ray spectra, including moderate absorption, 
for the range of the observation angles between $\theta = 30^\circ$ and $50^\circ$ (thick solid and dashed curves). The unabsorbed spectra are shown by the thin solid and dashed curves. The Lorentz factor of the blob II is equal to $\Gamma_{\rm II} = 2$  (for $\theta = 30^\circ$), and 1.5 (for $\theta = 50^\circ$).
The spectrum of the soft synchrotron radiation, emitted along the jet axis, is of the power law type with the spectral index $\alpha = 2$ in the energy range $\varepsilon_{\rm min} = 10^{-8}$ MeV and $\varepsilon_{\rm max} = 10^{-5}$ MeV. 
Its isotropic equivalent luminosity of the soft radiation is equal to $3\times 10^{45}$~erg~s$^{-1}$
for the small blob model and to $10^{46}$~erg~s$^{-1}$ in the large blob model. The spectrum of electrons (in the blob frame) is also of the power law type, with the spectral index linked to the spectral index of the soft radiation by $\beta = 2\alpha - 1$, above $E_{\rm min} = 2$ MeV. 
In the large blob model, the electron spectrum extends to $E_{\rm max} = 300$ GeV and $500$ GeV for the angles 
$\theta = 30^\circ$ and $50^\circ$ and in the small blob model up to $E_{\rm max} = 10$ TeV. The required energy in relativistic electrons is reported in Table~1. 
The radius of the blob II is fixed on $R = 3\times 10^{15}$~cm and the distance from the base of the jet on 
$z = 3\times 10^{16}$~cm for the small blob model and on $R = 3\times 10^{16}$~cm and on $z = 3\times 10^{17}$~cm for the large blob model.
For the comparison, we also show the $\gamma$-ray spectra from the SSC process occurring in the blob II, which are produced by this same population of electrons for these same parameters of the blob II as assumed in the case of the external blob model and for the angle $\theta = 30^\circ$ in the case of the small blob II model (dot-dashed curve) and for $\theta = 30^\circ$ (dot-dashed) and $50^\circ$ (dot-dot-dashed) in the case of the large blob II model.
The experimental results in the GeV-TeV $\gamma$-ray energy range are obtained simultaneously during the campaign between August 2010 and February 2011 with the ${\it Fermi}$-LAT and the MAGIC telescopes 
(see Aleksi\'c et al.~2014a).}
\label{fig7}
\end{figure*}

We propose an external blob radiation model for the high energy emission from the FR I radio galaxies which are supposed to be viewed by the observer at large angle in respect to the jet propagation. In this model, the isotropically distributed relativistic electrons in the outer blob II up-scatter mono-directional soft radiation produced in the inner blob I. As a result of such specific scattering geometry (see for details Aharonian \& Atoyan~1981), the angular distribution of the $\gamma$-ray photons does not peak at the jet axis. This is in contrast to the predictions of the homogeneous synchrotron self-Compton (HSSC) single blob models in which case electrons and soft synchrotron photons are isotropic in the blob reference frame.
In the HSSC model the strongest radiation is emitted along the jet axis due to the dominant Doppler boosting effect. On the other hand, considered here geometric model predicts the strongest radiation at some angle to the jet axis which depends not only on the Doppler boosting effect but also on the angular distribution of high energy emission in the blob II reference frame (see Figs.~3 and~4). We investigate the basic features of the high energy radiation as a function of the parameters describing the spectra of soft photons and relativistic electrons, the Lorentz factor of the blob II, and the location of the external observer in respect to the jet (Fig.~5). We conclude that the calculated $\gamma$-ray spectra depend in a systematic way on the parameters describing the spectra of soft radiation and relativistic electrons, but depend in a non-systematic way on the observation angle of the jet and the Lorentz factor of the blob II. 
 
We show that such a model can naturally explain the high energy emission observed from radio galaxies which are expected to be observed at large inclination angle to the jet axis. As an example, we interpret the emission from the FR I radio galaxy, NGC 1275, which is supposed to the viewed at the angle in the range $\sim$($30^\circ-50^\circ$) (Fig.~6). We conclude that the spectrum observed simultaneously by the {\it Fermi}-LAT and the MAGIC telescopes, in the GeV-TeV energy range, can be well described for the reasonable power in the low energy soft photons emitted from the blob I, equal to the isotropic equivalent synchrotron power of the order of a few $10^{45}$~erg~s$^{-1}$. Such emission power is within the range 
of the powers (a few $\sim$10$^{45}-10^{46}$ erg~s$^{-1}$) observed from the BL Lac type objects, e.g. Mrk 501 (Ahnen et al.~2017), Mrk 421 (Aleksi\'c et al.~2012b) or PKS 2155-304 (Aharonian et al.~2009, Abramowski et al.~2012). In the case of a relatively small optically thick blob ($R = 3\times 10^{15}$ cm), the total energy in the relativistic electrons in the blob II, able to explain the $\gamma$-ray emission from NGC~1275, should be equal to $\sim 10^{48}$ ergs, for the viewing angles not far from $\sim 30^\circ$.  However, the absorption effects for the small blob II viewed at the angle $50^\circ$ are too strong to provide correct description of the NGC 1275 spectrum. In the large blob II model, the absorption effects are small enough to provide correct description of the observed spectrum for the whole range of the observation angles $30^\circ-50^\circ$. However, in this case the energy in relativistic electrons should be much larger, i.e. of the order of a few $10^{49}$ erg, which is still acceptable in the case of blobs in jets of AGNs.

Finally, we note that proposed here model shows general similarities to the external disk model developed
in Dermer, Schlickeiser \& Mastichiadis~(1992) and Dermer \& Schlickeiser~(1993).  
In Dermer and Collaborators model, the radiation from the accretion disk has to dominate over the radiation produced in the inner part of the jet (e.g. inner blob). Such situation is expected in the case of OVV type blazars as seen out of the axis. Our model is more suitable for the radio galaxies in which case the disk 
emission is typically on a low level. Then, the scattering of radiation, produced 
in the inner blob, by electrons in the outer blob can likely dominate.

We would like to thank the Referee for important comments and suggestions.
This work is supported by the grant through the Polish Narodowe Centrum Nauki No. 2014/15/B/\ ST9/04043.






\vskip 0.7truecm

\leftline{REFERENCES}

\vskip 0.4truecm
Abdo, A.A. et al. 2009a ApJ 707, 55

Abdo, A.A. et al. 2009b ApJ 699, 31

Abramowski, A. et al. 2012 A\&A 539 149A 

Acciari, V.A. et al. 2009 Science 325, 444

Aharonian, F.A., Atoyan A.M. 1981 Ap\&SS, 79, 321, 336

Aharonian, F. et al. 2003 A\&A 403, L1

Aharonian, F. et al. 2006 ApJL 695, L40

Aharonian, F. et al. 2007 ApJL 664, L71

Aharonian, F. et al. 2009 A\&A 502, 749

Ahnen, M.L. et al. 2017, A\&A 603, 31A

Ait Benkhali, F., Chakraborty, N., Rieger, F.M. 2018 A\&A, submitted (arXiv:1802.03103)

Aleksi\'c, J. et al. 2012a A\&A 539, L2

Aleksi\'c, J. et al. 2012b A\&A 542, 100A

Aleksi\'c, J. et al. 2014a A\&A 564, A5

Aleksi\'c, J. et al. 2014b Science, 346, 1080A

Albert, J. et al. 2007 ApJ 669, 862

Albert, J. et al. 2008 ApJL 685, L23

Aliu, E. et al. 2012 ApJ 746, 141

Ansoldi, S., Antonelli, L.A., Arcaro, C. et al. 2018, A\&A, accepted (arXiv:1806.01559)

Baghmanyan, V., Gasparyan, S., Sahakyan, N. 2017 ApJ 848, 111

Barkov, M.V., Aharonian, F.A., Bosch-Ramon, V. 2010 ApJ 724, 1517

Bednarek, W., Protheroe, R.J. 1997 MNRAS 287, L9

Bednarek, W., Protheroe, R.J. 1999 MNRAS 310, 577

Begelman, M.C., Fabian, A.C., Rees, M.J. 2008 MNRAS 384, L19

Biretta, J.A., Parks, W.B.S., Acchetto, F.M. 1999 ApJ 520, 621

Blandford, R.D., McKee, C.F. 1976 PhFl 19, 1130   

Brown, A.M., Adams, J. 2011 MNRAS 413, 755

Brown, A.M.  BAhm, C., Graham, J., Lacroix, T., Chadwick, P., Silk, J. 2017 PRD 95, 063018

Chiaberge, M., Capetti, A. Celotti, A. 2001 MNRAS 324, L33

Dermer, C.D., Schlickeiser, R., Mastichiadis, A. 1992 A\&A 256, L27

Dermer, C.D., Schlickeiser, R. 1993 ApJ 416, 484

Dermer, C.D., Sturner, S.J., Schlickeiser, R. 1997 ApJS 109, 103

Dondi, L., Ghisellini, G. 1995 MNRAS 273, 583

Fujita, Y., Nagai, H. 2017 MNRAS 465, 94

Georganopoulos, M., Perlman, E.S., Kazanas, D. 2005 ApJ, 634, L33

Giannios, D., Uzdensky, D.A., Begelman, M.C. 2009 MNRAS 395, L29

Kataoka, J., Stawarz, \L ., Cheung, C. C. et al. 2010 ApJ 715, 554

Levinson, A., Rieger, F. 2011 ApJ 730, 123

Lucarelli, F. et al. 2017, ATel 9934

Mirzoyan R. 2017, ATel 9929

Moderski, R., Sikora, M., Coppi, P.S., Aharonian, F. 2005 MNRAS 363, 954 

Mukherjee R. \& VERITAS Collaboration 2017, ATel 9931

M\"uller, C. et al. 2014, A\&A 569, 115

Neronov, A., Aharonian, F.A. 2007 ApJ 671, 85

Onori, F., Ricci, F., La Franca, F. et al. 2017 MNRAS 468, L97

Pian, E., Vacanti, G., Tagliaferri, G. et al. 1998 ApJ 492, L17

Pivato, G. \& Buson, S. 2015, ATel 8219

Rieger, F.M. 2017, in High Energy Gamma-Ray Astronomy, AIP Conf. Series (ed. F. A. Aharonian et al.), vol. 1792, 020008 (arXiv:1611.02986)

Rieger, F.M., Aharonian, F.A. 2008 A\&A 479, L5

Rieger, F.M., Mannheim, K. 2002, A\&A, 396, 833

Roustazadeh, P., B\"ottcher, M. 2011 ApJ 728, 134

Sahakian, N., Yang, R., Aharonian, F.A., Rieger, F.M. 2013 ApJL 770, L6

Sari, R. 1997 ApJ 489, L37

Scharw\"achter J., McGregor P.J., Dopita M.A., Beck, T.L. 2013, MNRAS, 429, 2315

Sitarek, J., Bednarek, W. 2010 MNRAS 401, 1983

Tanada, K., Kataoka, J., Arimoto, M., Akita, M., Cheung, C.C., Digel, S.W., Fukazawa, Y. 2018 ApJ, in press (arXiv:1805.02361)

Tavecchio, F., Ghiselini, G. 2008 MNRAS 385, L98

Tingay, S.J., Preston, R.A., Jauncey, D.L. 2001 AJ 122, 1697

Vermeulen, R.C., Readhead, A.C.S., Backer, D.C. 1994 ApJ 430, 41

Wilman, R.J., Edge, A.C., Johnstone, R.M. 2005 MNRAS 359, 755

\end{document}